\documentclass[a4paper,twocolumn]{aa}
\usepackage{graphicx}
\newcommand\href[2]{#1}
\usepackage{amsmath}
\usepackage{wrapfig}
\usepackage[utf8]{inputenc}

\usepackage{etoolbox} 
\usepackage{breqn} 

\usepackage[switch]{lineno} 
\BeforeBeginEnvironment{dmath}{\begin{nolinenumbers}}
\AfterEndEnvironment{dmath}{\end{nolinenumbers}}
 \newcommand*\patchAmsMathEnvironmentForLineno[1]{
   \expandafter\let\csname old#1\expandafter\endcsname\csname #1\endcsname
   \expandafter\let\csname oldend#1\expandafter\endcsname\csname end#1\endcsname
   \renewenvironment{#1}
      {\linenomath\csname old#1\endcsname}
      {\csname oldend#1\endcsname\endlinenomath}}
 \newcommand*\patchBothAmsMathEnvironmentsForLineno[1]{
   \patchAmsMathEnvironmentForLineno{#1}
      \patchAmsMathEnvironmentForLineno{#1*}
 }
 \AtBeginDocument{
 \patchBothAmsMathEnvironmentsForLineno{equation}
 \patchBothAmsMathEnvironmentsForLineno{align}
 \patchBothAmsMathEnvironmentsForLineno{flalign}
 \patchBothAmsMathEnvironmentsForLineno{alignat}
 \patchBothAmsMathEnvironmentsForLineno{gather}
 \patchBothAmsMathEnvironmentsForLineno{multline}
}

\def\arcsec{$^{\prime\prime}$}

\usepackage{color}

\newcommand{\LEt}[1]{}

\begin{document}
  \title{Correction of the brighter-fatter effect on the CCDs of Hyper Suprime-Cam}
\author{
   Pierre~Astier\inst{\ref{aff:LPNHE-CNRS}}
   \and
   Nicolas~Regnault\inst{\ref{aff:LPNHE-CNRS}}
}
\institute{
  {LPNHE, (CNRS/IN2P3, Sorbonne Université, Université Paris Cité), 
  Laboratoire de Physique Nucléaire et de Hautes Énergies,
  F-75005, Paris, France}\label{aff:LPNHE-CNRS}
} 
\titlerunning{BF on HSC} \offprints{pierre.astier@in2p3.fr}
\date{Received Mont DD, YYYY; accepted Mont DD, YYYY}

\abstract{The brighter-fatter effect affects all CCD sensors to
  various degrees. Deep-depleted thick sensors are seriously affected
  and the measurement of galaxy shapes for cosmic shear measurements
  requires an accurate correction of the effect in science images. We
  describe the whole correction chain we have implemented for the CCDs
  of the Hyper Suprime-Cam imager on the Subaru Telescope. We derive
  non linearity corrections from a new sequence of flat field images,
  and measure their statistics, namely their two-pixel function.
  We constrain an electrostatic model
  from flat field statistics that we use to correct science
  images. We find evidence that some fraction of the observed variance and some
  covariances is not due to the combination of Poisson statistics and
  electrostatics -- and the cause remains elusive.
  We then have to ignore some measurements when deriving the electrostatic model.
  Over a wide range of image qualities and in the 5 bands of
  the imager, stars in corrected science images exhibit size variations
  with flux small enough to predict the point spread function for faint objects to an
  accuracy better than
  $10^{-3}$ for the trace of second moments -- and even better for the
  ellipticity and the fourth radial moment. This performance is sufficient
  for upcoming large-scale cosmic shear surveys such as Rubin/LSST.} 

\maketitle

\section{Introduction}
The brighter-fatter (BF) effect refers to a dynamical image distortion that
affects CCD sensors. The most spectacular manifestation of the effect
is that bright stars appear slightly bigger in size than faint ones, a
manifestation that is reflected the very name of the effect.
All studies of the effect have attributed it to distortions of the drift electric field
sourced by the charges stored in the pixel potential wells during image
integration. This modifies the apparent shape of bright objects and
the two-point statistics of uniform exposures, and thick CCDs are more
vulnerable to the effect than thinner sensors. Evidence of the effect
and the physical explanation can be found in \citet[and
references therein, G15 hereafter]{Guyonnet15}, together with the relation of the
effect with non-trivial flat field statistics. Electrostatic calculations
are shown to reproduce the data in \cite{Rasmussen16} and in \cite{Lage21} for
a specific sensor. 

On deep-depleted thick CCDs, bright stars thus generally appear bigger by a few
percent than faint stars, compromising the modeling of
the image point spread function (PSF) at a level that is not tolerable for large-scale cosmic
shear measurements (see e.g.,  \citealt{Mandelbaum-HSC}). The correction
method proposed in G15 relies on flat field statistics to constrain
the correction applied to science images, which mostly consists of 
correcting the recorded image for dynamically displaced pixel boundaries. The
method has been implemented for DECam in \cite{Gruen-PACCD-15}, and
for Hyper Suprime-Cam (HSC) on the Subaru telescope in
\citet[C18 hereafter]{Coulton-18} with some minor differences with
respect to G15.

The method proposed in G15 relies on first-order perturbations both in
the modeling of flat field correlations and when correcting science
images. In \citet[A19 hereafter]{Astier19}, the relation between
pixel area alterations and flat field statistics is extended to higher
orders, which removes significant biases from the analysis. In the
same paper, correcting for non-linearity of the video chain is
shown to play a potentially important role when constraining the BF
effect from flat fields. One other evolution since the G15 proposal
is that detailed electrostatic calculations have been shown to reproduce the
measured flat-field statistics, when the mandatory
manufacturing data is available (see, in particular, \citealt{Lage21}).
However, the CCD vendors do not necessarily release this data, or they do
not even have it available to the required level of accuracy. As we show in this paper,
there is also some detectable demographic variability among the CCDs of
the HSC camera, which are all of a unique type from a single vendor.
\cite{Gruen-PACCD-15} also detected some variability among DECam CCDs.
Thus, constraining the image corrections from measurements of the actual
sensors is still in order. 

In the present paper, we revisit the BF correction for the HSC camera
described in C18 and applied to science images in
\cite{Mandelbaum-HSC}. We take advantage of a new flat-field sequence
that allows us to re-determine both the non-linearity correction and
the two-point correlation function of flat fields. We apply to these
images some potentially important corrections (described in A19)
and we fit a variance and covariance model that has been improved since C18. We also
propose a different approach for transforming the information
extracted from flat fields into the correction of science images.
Finally, we test the correction of the images separately over
a broad range of image qualities and in the five bands of the camera. 

The flow of the paper is as follows. We first detail in
\S~\ref{sec:nonlin} why it is necessary to correct non-linearities prior to
measuring flat field statistics and the non-linearity measurement
itself.  In \S \ref{sec:ff_measurements}, we describe the measurements
of flat field statistics and the fit of the measurements, as well as the variability observed among the sensors. The information
extracted from flat field statistics is fundamentally insufficient to
correct the science images, thus, we describe in
\S~\ref{sec:sec:electrostatic_model} the electrostatic model we
use to derive the correction from the flat-field results and the
outcome of different fits we perform. Once we obtain models that allow
us to correct science images, we apply those to real data, as described in \S~\ref{sec:sci_corr}, and we compare the various outcomes. We face
the evidence that the BF correction is inadequate for the $y$ band and
we compute a physically motivated reduced correction for this band, which
we eventually apply to the science data. In \S~\ref{sec:discussion}, we
compute some PSF modeling quality indicators commonly used in the
context of shear estimation and conclude that the quality of our
correction exceeds the requirements for a large-scale cosmic shear
survey such as Rubin/LSST. It also fulfills the less stringent
requirements of the PSF fidelity implied by photometry accuracy of
high-redshift faint supernovae in order to estimate their distances,
which is our initial motivation for this work.

\section{Non-linearity correction}
\label{sec:nonlin}
\subsection{Importance of the non-linearity correction}
Following A19, we assume that the effective area A of a pixel
is linearly altered by the charge content of the sensor:
\begin{equation}
  \delta A = A \ g\ \sum_{i,j} a_{ij}  Q_{ij} , 
  \label{eq:delta_a}
\end{equation}
where $a_{ij}$ is a characteristic of the sensor and $Q_{ij}$ denotes
the charge content of the image, which evolves as light integration
goes on. The indices $i$ and $j$ refer to distances in pixel units
along the serial and parallel directions, respectively. In the same coordinates, $\delta A$ applies to the pixel located at $(i=0, j=0)$. Conventionally,
$a_{ij}$ is expressed in el$^{-1}$, $Q_{ij}$ in ADU and $g$ is
the gain in el/ADU. We note that, equivalently, we may set $g=1$ and express $Q$ in
electrons. From parity symmetry, we assume that: 
\begin{equation}
  a_{ij} = a_{|i||j|},
\end{equation}
so that we measure the $a_{ij}$ on the $i,j\geqslant 0$ quadrant.
If we consider a uniform image, with all $Q_{ij}$
identical, then $\delta A$ has to be zero, because of translation
symmetry. This imposes the following ``sum rule'': 
\begin{equation}
  \sum_{-\infty<i,j<+\infty} a_{ij} =0,
  \label{eq:sum_rule}
\end{equation}
where the sum extends to the four quadrants. This sum rule can also be
regarded as a consequence of area conservation. Since $a_{ij}$ is the
fractional pixel area change for a unit source charge, we refer
to these quantities as ``area coefficients.'' Since same-sign charges
repel each other, a pixel shrinks as it fills up, and
the self-interaction coefficient $a_{00}$ is negative. For
regular CCD operating conditions, all the other coefficients turn out
to be positive. The sum rule hence indicates that $|a_{00}|$ is much
larger (in absolute value) than any other coefficient.

Following Eq. \ref{eq:delta_a}, the expression of pixel
covariances in uniform exposures reads (see the derivation in A19):
\begin{dmath} 
  C_{ij}(\mu) =  \frac{\mu}{g} \left[\delta_{i0} \delta_{j0} + a_{ij} \mu g + \frac{2}{3} [\boldsymbol{a} \otimes \boldsymbol{a}]_{ij} (\mu g)^2
    + \frac{1}{3}[\boldsymbol{a} \otimes \boldsymbol{a} \otimes \boldsymbol{a}]_{ij} (\mu g)^3 + \cdots \right] + n_{ij}/g^2 \label{eq:C_ij} ,
\end{dmath}
where $\mu$ is the average (in ADU) of the uniform exposure. $n_{ij}$
refers to noises (expressed in el$^2$), and $n_{00}$ refers to the
usual read noise. From this expression, the relation between the
variance of uniform exposures and their average (usually called the
photon transfer curve or PTC) can be expressed as:
\begin{equation}
  C_{00} = n_{00}/g^2 + \mu/g + a_{00}\mu^2 + O(\mu^3),
  \label{eq:C00}
\end{equation}
where the first term is the read noise, the second is the Poisson
term, and the quadratic term is the first and largest contribution of
the BF effect: the slope of the PTC decays with the signal level (because
$a_{00}<0$). The fact that the PTC of CCDs grows less rapidly than the
signal level was initially noted in \cite{Downing06}, who also noted
that thanks to positive covariances, grouping data into bigger
pixels tends to restore the Poisson behavior. These non-trivial
flat-field statistics and the brighter-fatter effect per se were
unified under the same physics in \cite{Antilogus14}.

The analog readout chain can be affected by non-linearity and
a common distortion is a quadratic contribution:
\begin{equation}
\mu_m = \mu_t + k \mu_t^2,
\end{equation}
where $\mu_m$ and $\mu_t$ refer to measured and true values, and $k$ is
a small quantity describing the distortion. The last term can
equivalently refer to either value of $\mu$. For the variances, we have:
\begin{align}
  V_m & \simeq V_t \left(\frac{\partial \mu_m} {\partial \mu_t} \right )^2 \nonumber \\
      & = V_t (1+2 k\mu)^2 \nonumber \\
  &\simeq V_t (1+4k\mu)  \nonumber \\
  & =  (n_{00}/g^2+\mu_t/g+a_{00}\mu_t^2)(1+4k\mu_t) \nonumber \\
  & =  n_{00}/g^2+\mu_m/g +(a_{00}+ 3k/g)\mu_m^2 + O(\mu^3). \label{eq:bias_a00}
\end{align}
Thus, comparing Eqs. \ref{eq:bias_a00} and \ref{eq:C00}, we find that a quadratic non-linearity, if it is not corrected for, biases the
measurement of $a_{00}$, which is the largest coefficient describing the
BF effect. The same calculation applied 
to covariances shows that there is no effect at the $\mu^2$ level, hence the measurement of the other $a_{ij}$ coefficients is much
less sensitive to a quadratic non-linearity than $a_{00}$.
\begin{figure}[ht]
  \begin{center}
    
\includegraphics[width=\linewidth]{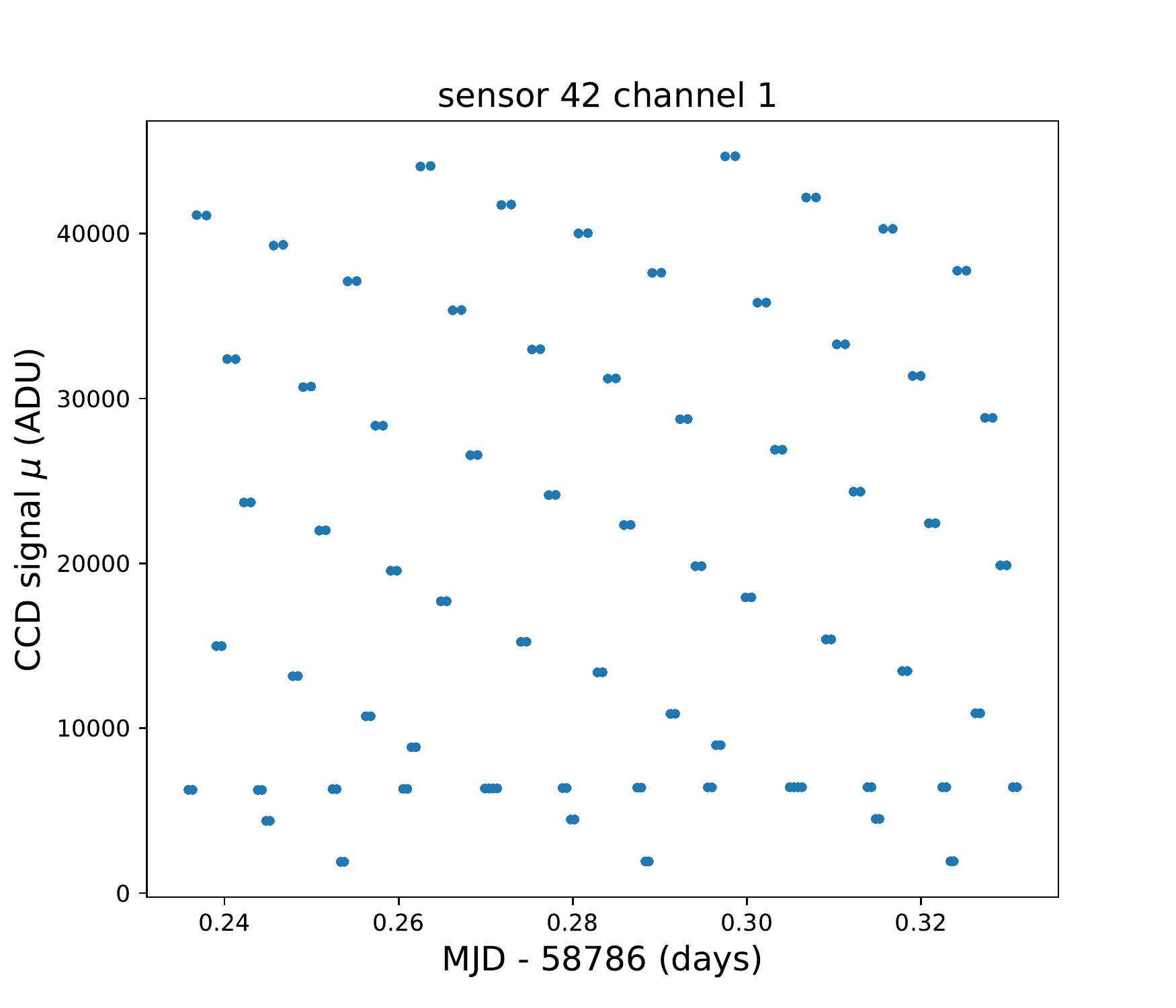}
\caption{CCD signal values as a function of time for a single channel
  during a sequence of uniform exposures illuminated by a lamp in the dome. 
  The intensities are deliberately not ordered in a monotonic
  way, so that the drift of the illumination system can be separated from
  the non-linearity. The sequence was designed by N. Yasuda (IPMU). 
  \label{fig:val_vs_time}}
\end{center}
\end{figure}

\begin{figure}[ht]
  \begin{center}
    
\includegraphics[width=\linewidth]{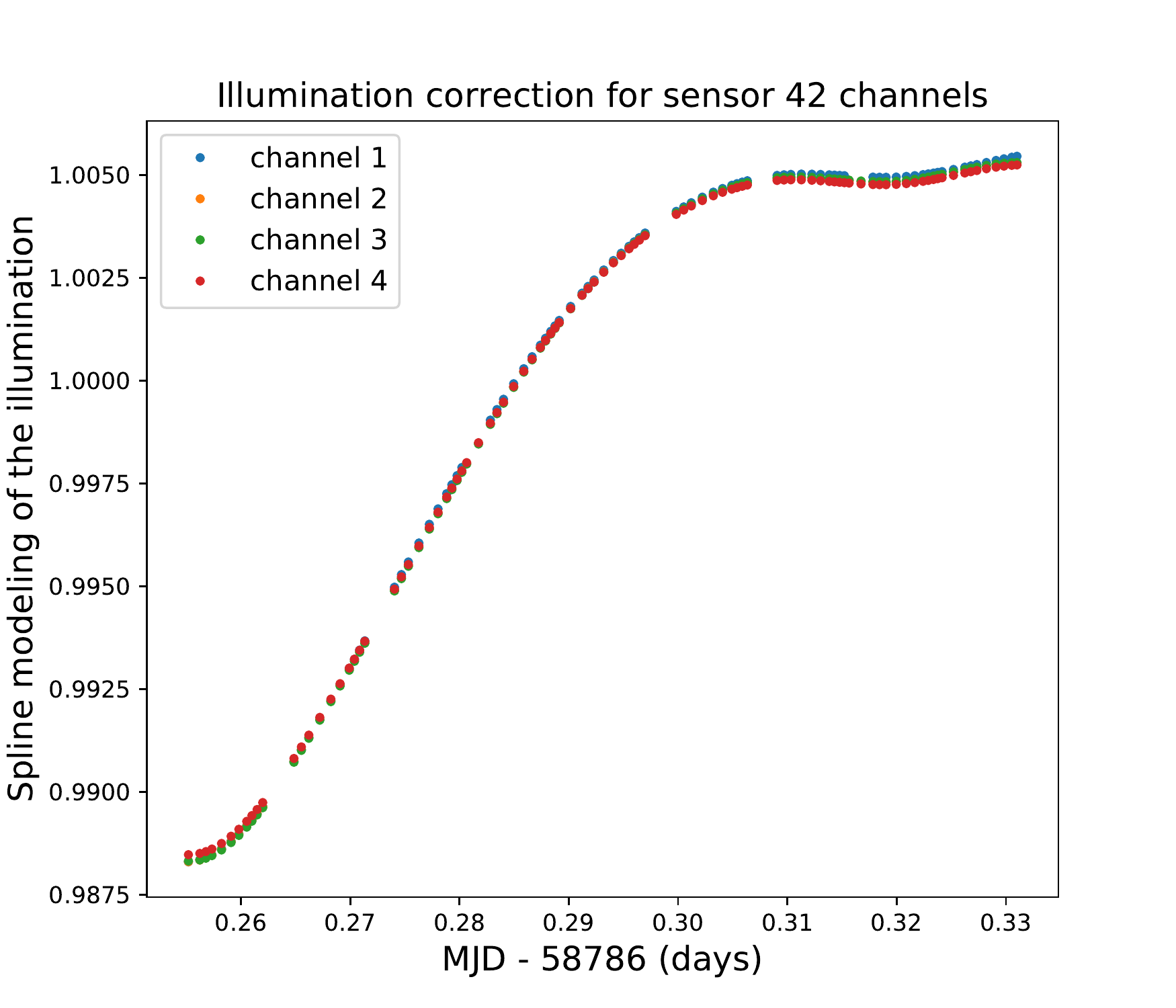}
\caption{Spline correction from the minimization of Eq. \ref{eq:non_lin_chi2} for four different channels, normalized to its average.  Since this is presumably due to lamp variations, it should be similar for all channels. The maximum difference is in the  $10^{-4}$ range.
  \label{fig:spline_corr}}
\end{center}
\end{figure}

\subsection{Measurement of response non-linearity}
\begin{figure}[ht]
  \begin{center}
    
\includegraphics[width=\linewidth]{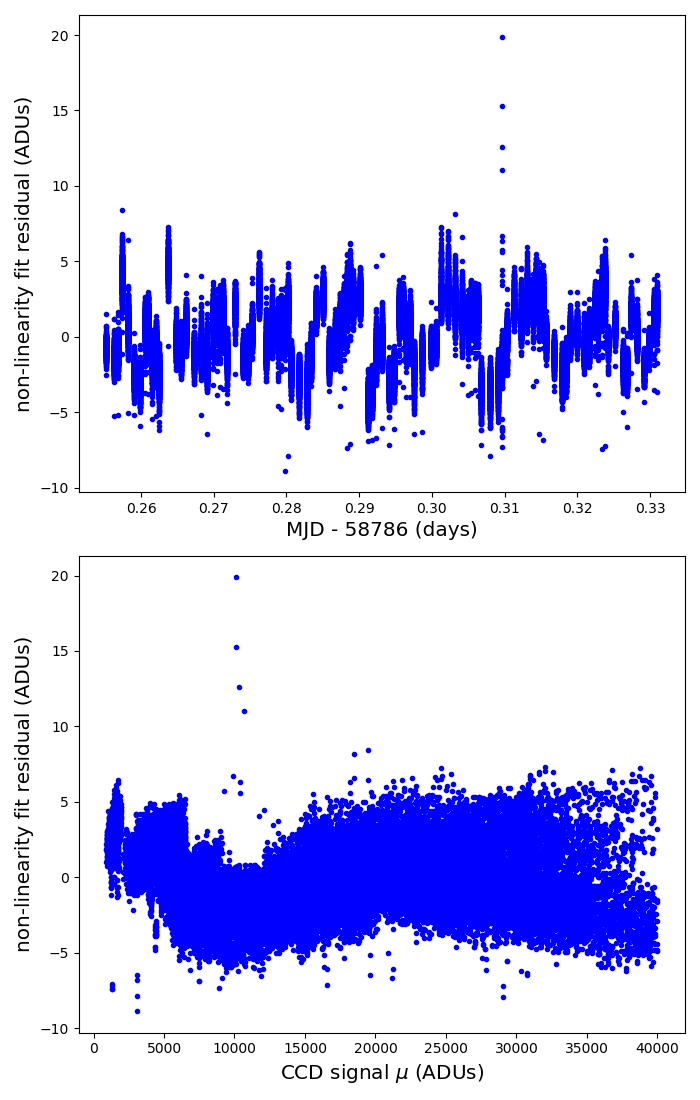}
\caption{Fit residuals as a function of date (top) and signal
  level (bottom) for all channels of all sensors. There are no obvious
  needs for a more sophisticated
  model as a function of date nor signal level. The rms residual is
  about $3\ 10^{-4} \mu$.
  \label{fig:plot_res}}
\end{center}
\end{figure}

\begin{figure}[ht]
  \begin{center}
    
\includegraphics[width=\linewidth]{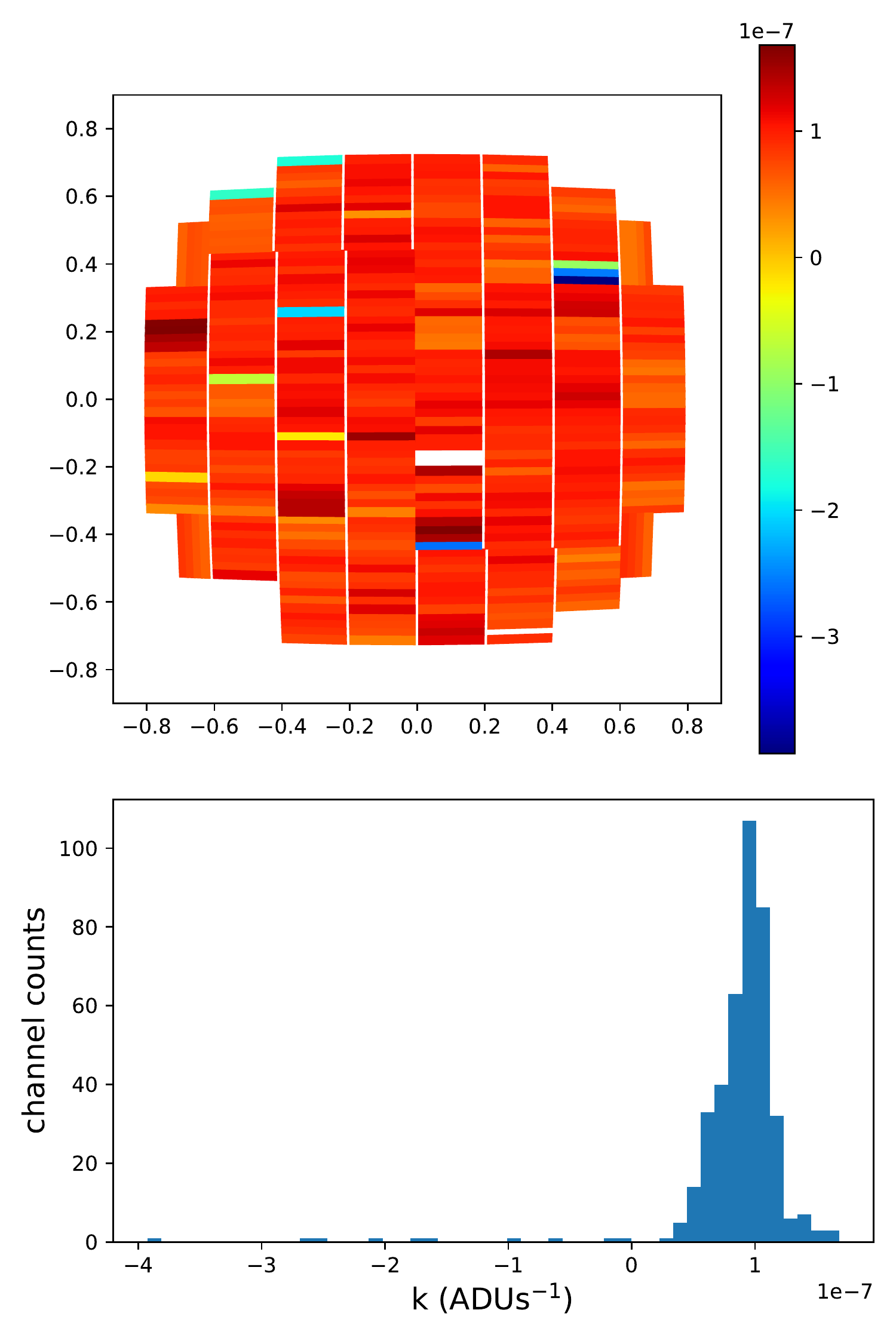}
\caption{Values of measured quadratic non linearity $k$ (in
  Eq. \ref{eq:non_lin_chi2}) in the focal plane (top) and their
  distribution (bottom). Most of the channels cluster around
  $10^{-7}$, which corresponds to a correction of $\sim 0.4$\% around
  maximum CCD signal.
  \label{fig:k_hist}}
\end{center}
\end{figure}
The Hyper Suprime-Cam instrument is installed at the prime focus of the
8.2-m Subaru telescope on the Mauna-Kea summit in Hawaii. A detailed
description of the instrument can be found in \cite{HSC18} and here we provide only the salient features for our work. The HSC camera
contains 104 science sensors, which are deep-depleted Hamamatsu
2k$\times$4k CCDs with four read-out channels on each sensor. The sensors
are 200~$\mu$m thick, the pixel size is 15 $\mu m$, which corresponds
to $0.185$\arcsec,\ on average, over the focal plane.  The saturation
of sensors occurs around 120,000 el and the video channels have a
typical gain of 3.2 el/ADU. All channels may a priori exhibit
different non-linearities.

In order to measure the non-linearity of the camera video chains, we
rely on a new sequence acquired on 2019/10/30, of uniform exposures
known as ``dome flats''  obtained at night by illuminating
a screen attached to the dome, toward which the telescope is pointed.
The
signal level is controlled by varying the exposure time. The
illumination system is not equipped with a photodiode or any similar
device and we use the exposure time as a proxy for the
amount of light received by the CCD. Obviously, this is an acceptable proxy
only if the illumination system is stable
over the whole sequence or if its variations can be modeled. In order
to disentangle a drift of the lamp intensity from a genuine
non-linearity, the various exposure times are not ordered time-wise in
a monotonic way, as shown in Fig. \ref{fig:val_vs_time}.

For a given video channel, we collect from every exposure (labeled
$i$) an average signal, $\mu_i$ (in ADUs), an exposure time, $T_i$, and
the (Julian) date at which it was acquired, $d_i$. We minimize the quantity:
\begin{equation}
  \sum_i \left( (T_i+O)S(d_i) -\mu_i -k\mu_i^2 \right )^2,
\label{eq:non_lin_chi2}  
\end{equation}
where $S$ represents a seven-knot spline meant to describe the lamp
variation with time, $O$ is a potential systematic shutter offset,
$k$ represents
the quadratic non-linearity, and $T_i$ is the required exposure time.
The shutter offset is meant to account for a (small) difference
between the required and realized exposure times. We do not find 
any compelling evidence for a global shutter offset, nor a spatial
trend in the focal plane, that could arise from such a large blade
shutter. Hence, we set $O=0$. Since the illumination may evolve
along the sequence in slightly different ways across the focal plane,
we do not enforce the spline $S$ to be the same for all channels. We
eventually compare the outcome of different fits (on the same sensor)
in Fig. \ref{fig:spline_corr}. The fact that these corrections are
almost indistinguishable indicates that the non-linearity and the lamp
variations can be robustly separated from this data set.

Figure \ref{fig:plot_res} displays the fit residuals (from
Eq. \ref{eq:non_lin_chi2}) as a function of signal level or of date.
Those are globally small. A more flexible model as a function of date
would not reduce much these residuals, highly correlated over the channels
at the same date. We can attribute this scatter to random fluctuations of the
lamp or of the shutter. The fit residuals as a function of signal
level do not strongly call for a more flexible or more general non-linearity
model than a quadratic correction: the largest systematic residual is
found around 10,000 ADUs and amounts to a few ADUs. However, for some
yet unknown reason, we had to discard exposures before
$d_i-58786<0.255$ because they exhibit large residuals (tens of ADUs)
consistently over channels. One might attribute this to some erratic
behavior of the lamp that vanishes after about 0.5 h. While these first
exposures are rejected
from the non-linearity analysis, they are used to 
measure the flat field statistics in what follows.

In Fig. \ref{fig:k_hist}, we display the values of quadratic non-linearity
in the HSC focal plane in order to visualize the absence of geometrical
trend that could occur from a spatially variable evolution of the
illuminating system. The distribution peaks around $k = 10^{-7}$~ADU$^{-1}$,
which corresponds to a correction of $\sim
0.4$~\% at the maximum CCD signal. We go on to show soon that $a_{00}$
averages to $-1.3\ 10^{-6}$ and the channel gains are around 3; if one
ignores quadratic non-linearity (Eq. \ref{eq:bias_a00}), the bias
affecting $a_{00}$ amounts to $\sim$8\%. $a_{00}$ is the best measured area
coefficient and essentially drives the scale of the BF correction.
We might note that such a bias unavoidably causes a violation of the
sum rule and, hence, translating such a set of area coefficients
into a correction is somehow arbitrary because sensible image corrections
have to derive from areas changes that sum up to zero. 
We also note that that eliminating this trivial cause of sum rule
violation allows us to detect more subtle problems in \ref{sec:electro_fit}.

\section{Measurement and fit of flat-field statistics}
\label{sec:ff_measurements}
\subsection{Measurements}
Measuring the variances and covariances of flat fields allows one to
measure the dynamical alteration of pixel areas due to electrostatic
distortions; namely, we measure variances and covariances of
flat field exposures (which we may call the two-point function of the
exposure) at various signal levels, in order to measure the $a_{ij}$
coefficients (Eq. \ref{eq:delta_a}) by fitting Eq. \ref{eq:C_ij} to
the measurements. We closely follow the procedure described in A19: we
first correct for non-linearity (see the previous section) and correct
for deferred signals.

\begin{figure}[ht]
  \begin{center}
    
\includegraphics[width=\linewidth]{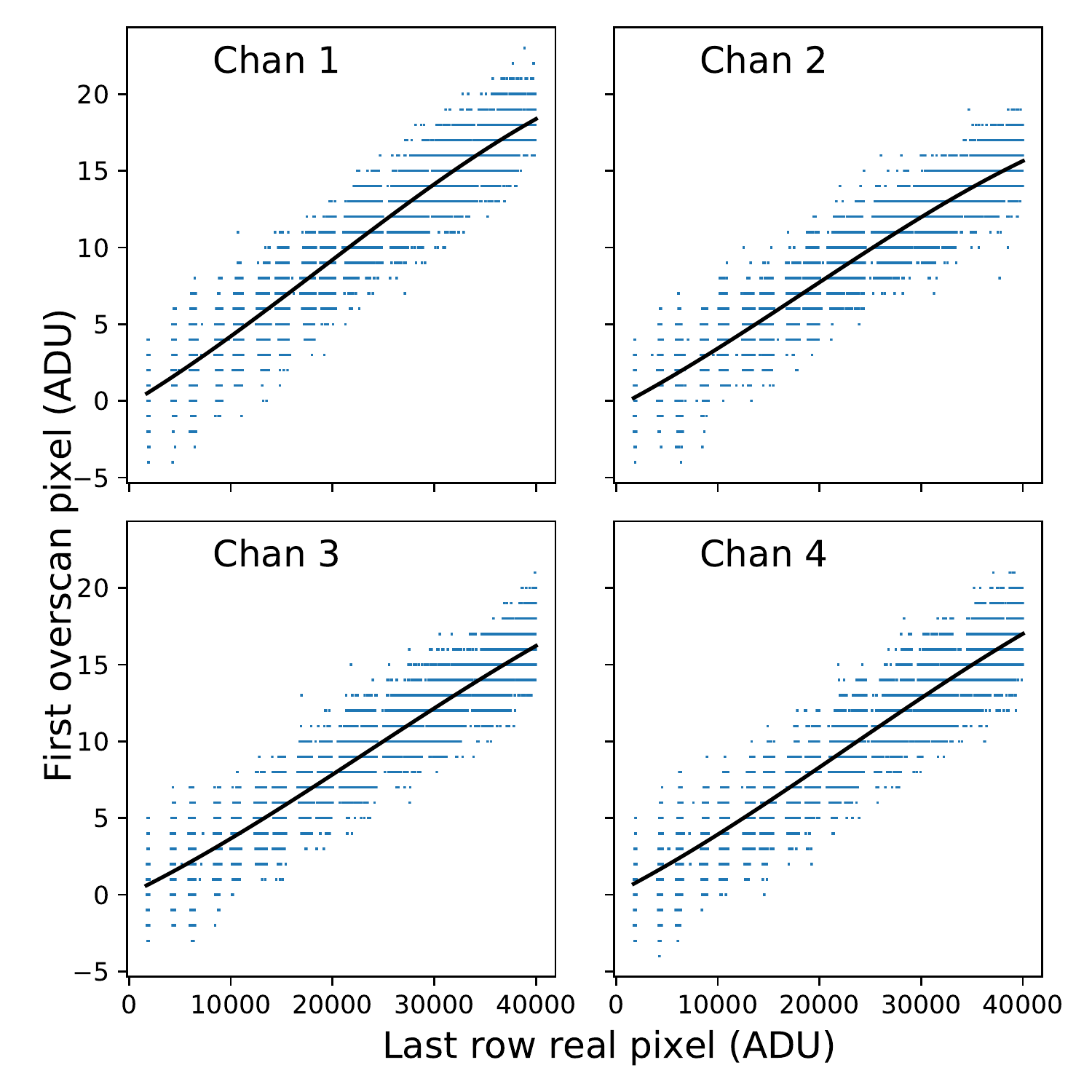}
\caption{Nearest serial neighbor deferred signals for the four channels of sensor 42.
  The values are small and vary almost linearly with the signal. The model
  continuous lines are the cubic polynomial model that we use for correction.
    \label{fig:cti}}
\end{center}
\end{figure}

A deferred signal is by definition collected after (in the serial time
chain) the pixel it belongs to. In order to measure those, we closely
follow the procedure described in A19, and find that there are no
significant deferred signals beyond the nearest serial neighbor.
We display a few examples of the nearest neighbor deferred signals in Fig. \ref {fig:cti}:
those are small and almost linear, indicating that they mostly originate 
from electronics. We tabulate one correction curve per channel, and for
correcting the effect, we typically ``place back'' less than 0.1~\% of
a pixel into the preceding one. Failing to carry out this correction results
in the covariance, $C_{10}$, exhibiting a linear contribution ($\propto \mu$),
which cannot be absorbed by the model of Eq. \ref{eq:C_ij}, hence biasing
$a_{10}$. For our CCDs, at the maximum signal level, the contribution of
uncorrected deferred signals to $C_{10}$ would represent about 25\% of the
electrostatic value. 

When computing image statistics, we mask any
outlier pixels detected in the images and use the pixel mask applied
to science images at the epoch of the sequence. These pixel masks flag
pixels that repeatedly depart from their neighbors in uniform
exposures and pixels on the sides of the focal plane which are
heavily vignetted (or fully blind). We account for the masked
pixels in covariance computations in a standard way, described in
Appendix A of A19. The bias of variance and covariances due to
masking outlier pixels from images themselves is compensated for
as described in \S 5.2 of A19.

We measure each video channel
independently and produce variance and covariance curves using the 66
flat pairs of the sequence. We measure the covariances on the difference
of flat pairs acquired at the same signal level to eliminate
contributions from illumination non-uniformity or sensor sensitivity
variations. These 66 flat pairs were acquired in $g$ band. Contrarily
to C18, we carry out the measurements on all HSC science sensors.

\subsection{Fits of the covariance curves}
\begin{figure}[ht]
  \begin{center}
    
\includegraphics[width=\linewidth]{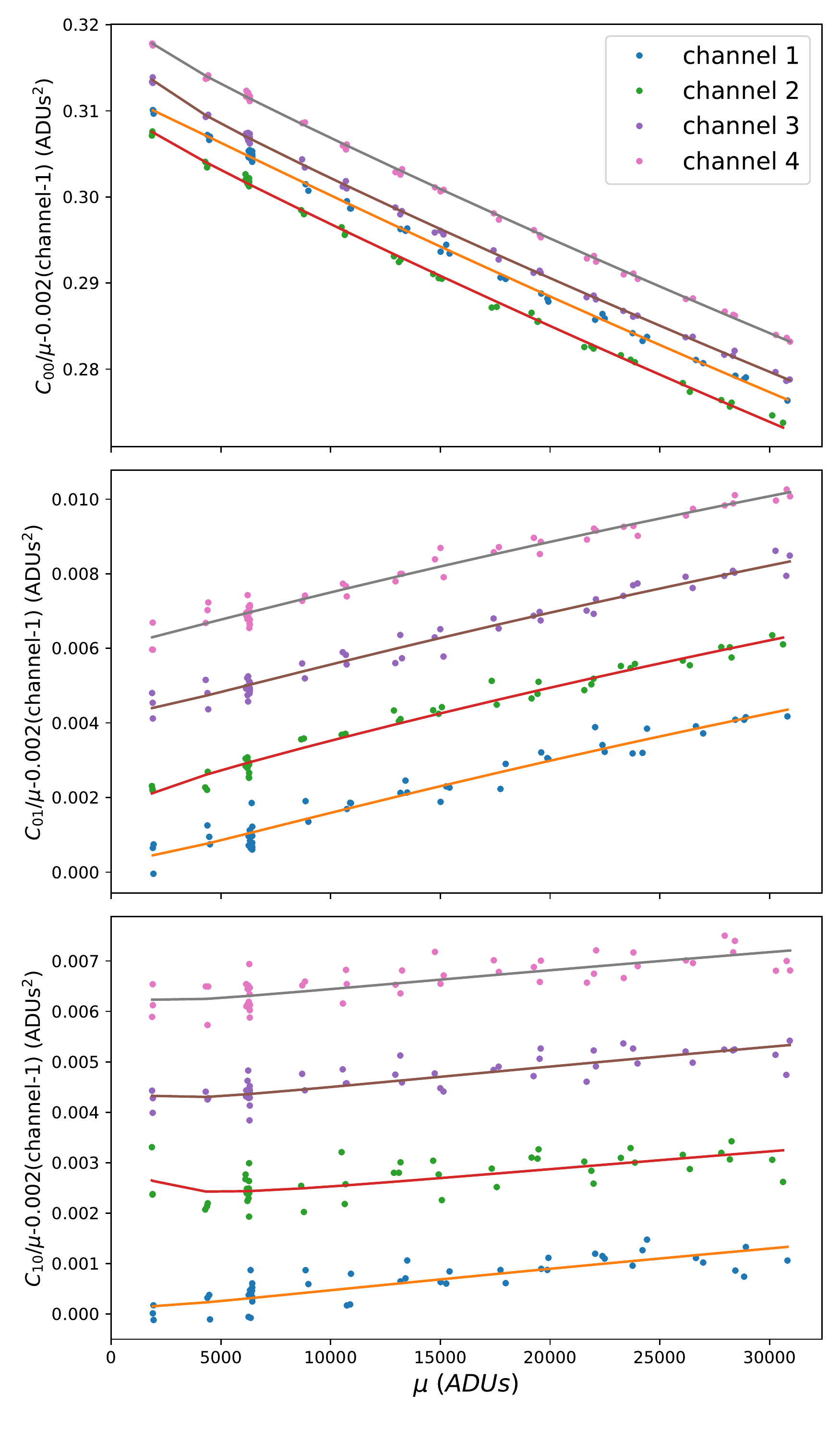}
\caption{Measurement and fits of the variance ($C_{00}$) and the two
  nearest neighbor covariances ($C_{10}$ and $C_{01}$) for the four
  channels of sensor 42. The channels have been offset for visual
  convenience. The decay of $C_{00}/\mu$ with $\mu$, originally noted in
  \cite{Downing06}, is the most obvious effect of the BF
  effect. The spread of measurements is compatible with shot noise.
  \label{fig:ptc_and_co}}
\end{center}
\end{figure}

We then fit the measurements using Eq. \ref{eq:C_ij}, with
weights derived from shot noise. This equation contains terms up to
$\mu^4$, and including one extra order would change the predictions by
a few $10^{-5}$, which is about 100 times less that the smallest shot noise.
Fig. \ref{fig:ptc_and_co} displays the measurements and fits for a few
$(i,j)$ pairs and for the four channels of a sensor. One can readily note
that the slopes for $C_{10}$ (serial neighbors) and $C_{01}$ (parallel neighbors) differ by a factor of more
than 4, which is due to the very different mechanisms that confine the
charges into the pixel wells for the serial and the parallel
directions: while the boundaries between columns are static and
implanted, the boundaries between rows result from the potentials
of the parallel clock stripes that are swung during image read-out
in order to move charges towards the serial register and the output amplifier.  The parallel
pixel boundaries are then somehow weaker than the serial ones and this
is reflected on the pixel covariances of uniform exposures. We perform
the measurements in the range $i,j<10$ and fit Eq.
\ref{eq:C_ij} to all the measurements of a given channel at once.

\begin{figure}[ht]
  \begin{center}
    
\includegraphics[width=\linewidth]{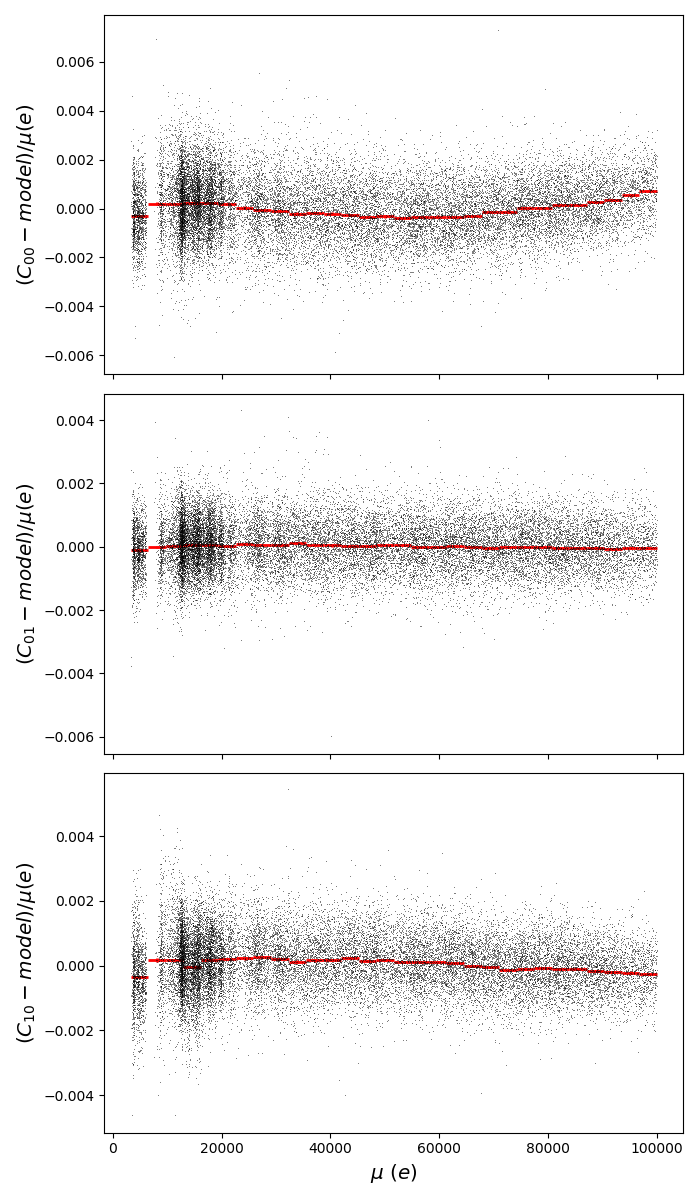}
\caption{Residuals of fits of variance and covariance measurements,
  all expressed in electrons using the gains from the fits, so that
  the data from different sensors can be averaged. The red signs are
  binned averages of the residuals from individual fits.
  \label{fig:plot_fit_res}}
\end{center}
\end{figure}

In order to question whether the fitted model (Eq. \ref{eq:C_ij})
properly describes the data, we plot in Fig. \ref{fig:plot_fit_res}
the fit residuals of the variance and the nearest covariances. Since
we are studying effects related to sensors, we turned the averages and
covariances into gain-free quantities by converting them into
electrons, using the gains obtained from the fits in the previous
paragraph. An examination of Eq. \ref{eq:C_ij} shows that the
coefficients of the $\mu^3$ terms are entirely determined by
combinations of the coefficients of the $\mu^2$ terms. The residuals
in Fig. \ref{fig:plot_fit_res} exhibit some curvature, which is compatible with
a mismatch between the coefficients of $\mu^2$ and $\mu^3$. For the
residuals of $C_{00}$ (top), the curvature of the residuals represents
about 15\% of the cubic term of Eq. \ref{eq:C_ij} (thus illustrating the importance of
these higher order terms). For $C_{01}$, the curvature represents
about 30\% of the cubic term, while they have similar values for the
residuals to $C_{10}$. We will propose later an explanation for the
curvature of residuals of $C_{00}$ and $C_{01}$, but this explanation
does not apply to $C_{10}$.

\subsection{Demography of HSC sensors}
\begin{figure*}[ht]
  \begin{center}
    
\includegraphics[width=\linewidth]{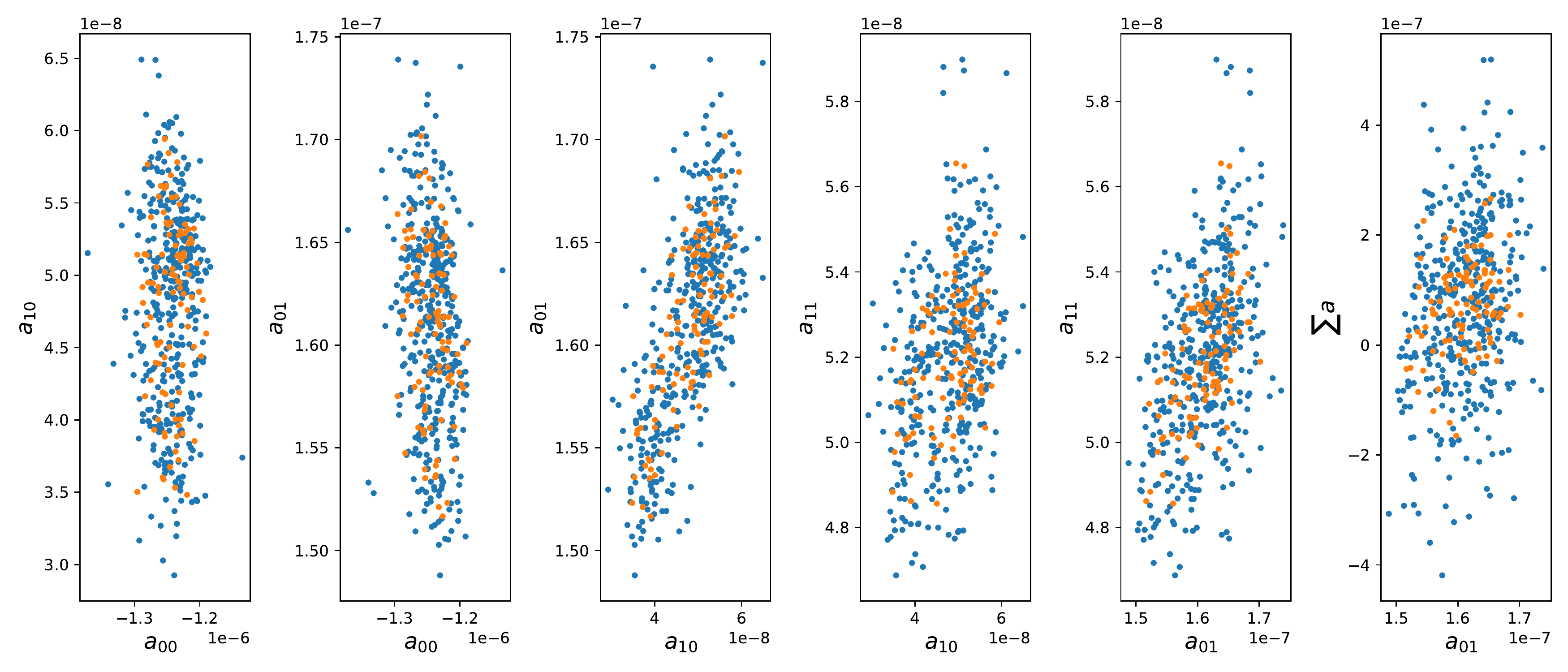}
\caption{Distribution of area coefficients over the $\sim$400 hundred
  channels of HSC (blue) and the averages over each sensor
  (orange). All $a_{ij}$ coefficients are expressed in el$^{-1}$. We
  may note that the trends for channels and sensors are similar, which
  indicates some homogeneity within sensors (except for $\sum a$, where the
  scatter is dominated by shot noise). $\sum a$ is the sum for
  $-10<i,j<10$.
  \label{fig:plot_a_dist}
}
\end{center}
\end{figure*}

We may now investigate the demography of channels and sensors
regarding the BF effect. In Fig. \ref{fig:plot_a_dist},
we display scatter plots of the well-measured $a$ coefficients. We
can see the correlations, which are similar at the channel and sensor
level, indicating that the variations are weaker within sensors. We
can note that $a_{00}$ does not correlate to the other
coefficients, and that the apparent correlations are in a direction
that does not preserve the ``sum rule.'' This may indicate that
they are not due to genuine variations of the BF effect
among the sensors.  We may also observe that on average the sum of $a$ is
positive:
\begin{equation}
\left< \sum_{-10<i,j<10} a_{ij} \right>_{channels}\simeq 7\ 10^{-8}.
\end{equation}
Since all $a_{ij}$ at larger distances are positive, the sum to
infinity is even farther from 0. Since $<a_{00}> \simeq -1.3
\ 10^{-6}$, the violation of the sum rule is hence at least 5\% of $a_{00}$ and we eventually see that it reaches $\sim$10\% once we have a model to evaluate the large-distance contributions. This
means that the sum of all covariances rises with signal level faster
than Poisson, at variance with expectations from statistics (see Eq. 8 in A19). Our
measurements are thus affected by some noise, which increases with
signal level and contributes as $\mu^2$.

Since the variations of well-measured $a_{ij}$ is at most 10 \% across
channels, we decide to average the measurements in order to average the
shot noise at large distance. Once equipped with this average, we fit
an electrostatic model to it that, by construction, satisfies the sum
rule. In the next section, we show that the mismatch between the
model and the data is indeed localized.

\section{Electrostatic fit}
\label{sec:sec:electrostatic_model}
\subsection{From the area coefficients to science image corrections} 
We study the BF effect in order to suppress its impacts from the
science images. The scheme proposed in G15 consists of
evaluating from the science image itself the motions of pixel boundaries
with respect to a perfect grid, as well as evaluating via
interpolation how much charge was flowing over these pixel boundaries
during integration, and then placing this amount of charge back where it
belongs.

We can readily note that the measurements we have performed so far
constrain the change in the area of a pixel, but do not tell how its
shape is altered. If we describe the shape change by different
motions of the serial and parallel sides of a
pixel, this simplistic shape description already requires two
quantities per pixel, while we only have one. Defining a shape change
by distinct serial and parallel boundary shifts means in turn that
we have to ``split'' the $a_{ij}$ coefficients (into $a_{ij}^N,
a_{ij}^W,a_{ij}^S,a_{ij}^E$) and rely on area-conserving symmetries
such as $a_{10}^W \equiv -a_{20}^E$. We can also regard $a_{ij}$ as the
discrete divergence of the pixel boundary displacement field and we
are faced with the ill-posed problem of determining a vector field
from its divergence, in a a 2d discrete space. 

Various implementations of the correction method proposed in
G15 differ in the way they perform this promotion of pixel
area change into two directional components. In G15 and in
\cite{Gruen-PACCD-15},  some ratios are imposed, which were estimated from the area
changes themselves. C18 argue that the motions of pixel
boundaries is a 2D vector field that results from the (perturbation)
electric field sourced by the charges that represent the image, and,
hence, the 2d displacement field is (as is the 3d electric field) curl-free.
The ``scalar'' $a_{ij}$ field can then be transformed into a 2d vector
curl-free field. While this curl-free assumption seems appealing, we show in \S~\ref{app:curl_free} that it turns out to be violated
by a solution of the Poisson equation. We may guess (and we show below) that the pixel boundaries motions are proportional to
the integral over pixel boundaries of the perturbating electric
field. These integrals would themselves be curl-free (as is the electric
field) if the integration paths were all identical for serial and
parallel boundaries.  The fundamental anisotropy (e.g., from the ratio of 
$a_{01}$ and $a_{10}$) indicates that this is not true at small distances, and questions the curl-free hypothesis applied to the pixel boundaries
displacement field. Even without this anisotropy, we might also question
whether the curl-free property of the continuous 3D electric field is
precisely transferred to a curl-like combination of finite differences
over the pixel lattice.

Our practical approach is to promote the discrete $a_{ij}$ scalar field into
a discrete 2D vector field relying on electrostatics; namely, to propose to fit the geometrical parameters of a simple electrostatic
model of the perturbating electric field to the data and to evaluate the
2D vector field we need from the model. We note that since
the model reports actual pixel areas, the sum rule is by
construction satisfied by the outcome of a fit.

\subsection{The electrostatic model}
\label{sec:elec-model}
We are interested in pixel boundary shifts under the influence of
stored charge and here we  sketch a first-order perturbation scheme
that allows us to predict these shifts from a few geometrical
quantities.  The first ingredient is the electric field sourced by the
stored charges inside the sensor, which we refer above as the
``perturbating'' electric field (because it adds to the drift field
that defines pixels). We assume that the field sourced by collected
charges causes no rearrangement of charge within the bulk of the device.
We model this perturbating field as sourced by a
charge between two infinite grounded equipotentials figuring the light
entrance window (on which the drift voltage is applied) and the
parallel clock stripes. This is a text-book problem and using
the image charge technique, we can write the
corresponding potential as:
\begin{align}
  \phi(\rho,z) = \frac{Q}{4\pi \epsilon} \sum_{n=-\infty}^{\infty}  &1/\sqrt{\rho^2+(2nt+z_q-z)^2} \nonumber \\
   & -1/\sqrt{\rho^2+(2nt-z_q-z)^2}, 
  \label{eq:potential}
\end{align}
where $\rho^2 = x^2+y^2$, $t$ is the sensor thickness, the source
charge $Q$ is located at $(0,0,z_q)$, and the equipotential planes are
at $z=0$ (representing the parallel clock stripes) and $z=t$ (the
light entrance face of the sensor). We can check that
$\phi(\rho,0)=\phi(\rho,t) = 0$: in these cases, each positive term
(first line) of the series has an exact opposite among the negative
terms (second line), which is how the image technique works. This expression
for the potential converges poorly with $n$ for $\rho$ much larger
than $t$ and alternative expressions should then be used
\citep{Pumplin69}. For the fit, we do not need to evaluate the model
at $\rho > t$.

\begin{figure}[ht]
  \begin{center}
        
\includegraphics[width=0.8\linewidth]{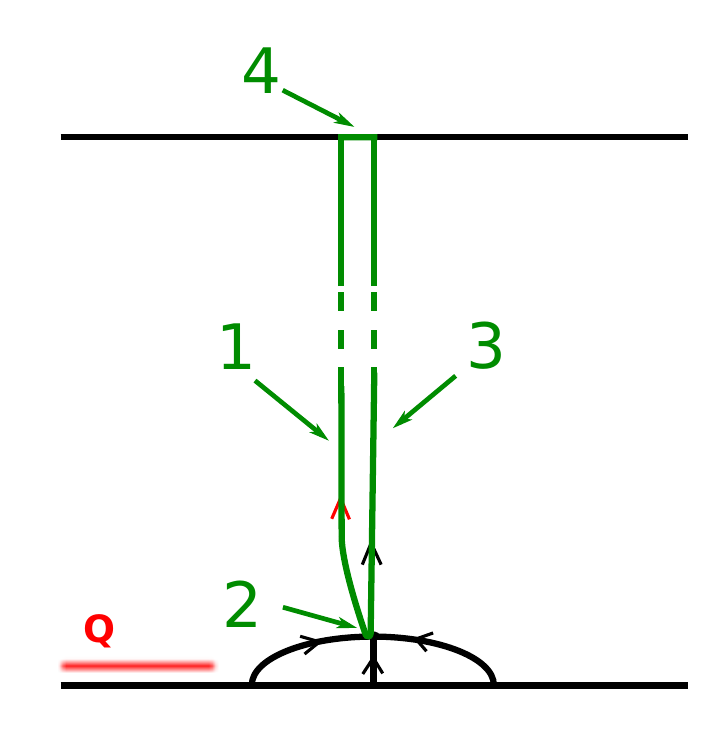}
\caption{Contour to which we are going to apply Gauss's theorem to
  relate the boundary shift (the length of segment 4) to the electric field
  sourced by the charge Q.
    \label{fig:fig-drift1}
}
\end{center}
\end{figure}
The field that results from the normal operation of the sensor drives
the charges into the pixel wells.  The drift lines on the pixel
boundaries have a peculiar point at which the field is null and the
potential has a saddle point. This point is commonly located a few
microns away from the clock stripes (at z=0). These points are located
at different heights above serial and parallel boundaries, and this
difference contributes to the anisotropy of flat field small-distance
covariances.  We display in Fig. \ref{fig:fig-drift1} a contour
crafted to relate the pixel boundary shifts to the
source charge using Gauss's theorem, applied to the total electric field,
the drift field plus the perturbation introduced by the charge.
This contour is made from four segments (numbers matching Fig.
\ref{fig:fig-drift1}). Segment 1 corresponds to the perturbed drift line that separates
  two pixels and the integral of the transverse electric field is null.
    Segment 2 represents the short path between the unperturbed
  zero-field point and the perturbed zero-field point and is very close to a drift line; the integral of the electric field is then a second order quantity, which we ignore.
Along segment 3, the integral of the unperturbed field vanishes  because it is the unperturbed drift line; 
we are thus left with the integral of the perturbation field.
Over segment 4, 
the integral of the field is the product of the drift field, $E_d$, times the boundary displacement, $d$, that we wish to estimate.

So, Gauss's theorem is finally expressed as:
\begin{equation}
  \int_{z=t}^{z=z_0} E_Q^T(x_b,y_b,z) dz + d E_d = \int_C \rho/\epsilon
  \label{eq:model0}
\end{equation}
where $z_0$ is the zero-field point altitude, $x_b$ and $y_b$ are the
coordinates of the unperturbed drift line, $E_Q^T$ is the field transverse to the
boundary sourced by the charge $Q$, $E_d$ is the drift field (at the
top), and the right-hand side is the bulk charge contained inside the
contour, because the Silicon material can contain residual impurities
sourcing a small bulk electric charge.
We do not know this right-hand side, but we can assume that
(to first order) it is proportional to $d$, our perturbation parameter. So,
we eventually obtain:
\begin{equation}
  d \propto \int_{z=t}^{z=z_0} E_Q^T(x_b,y_b,z) dz .
  \label{eq:model1}
\end{equation}
In other words, the boundary displacement is proportional to the integral of
the perturbating (transverse) field over the unperturbed trajectory.
This is usually called the Born approximation.
In a charge-free material, the
normalization is expressed as $1/E_d$. The expression assumes
that charges are produced when light enters the sensor, at $z=t$. For
the reddest bands, we should instead account for the conversion depth
of photons in the sensor, as we will do in \S
\ref{subsec:y-band-model}.  This expression should be averaged over
the pixel side in the direction perpendicular to the
Fig. \ref{fig:fig-drift1}.

In order to constrain the rhs of the expression, it is tempting
 to carry out the the measurements at different values of $E_d$
by changing the drift voltage applied to the entrance side of the sensor. 
However, changing the drift actually alters $E_d$ but also $z_0$ (for both flavors of
boundaries), so measuring the sensor under different drift voltages
would not help significantly at constraining the model. We may, in
principle, predict the proportionality coefficient of Eq.
\ref{eq:model1} because the (empty CCD) charge density, thickness,
applied drift voltage, and drift field are related by basic
electrostatics. We nonetheless stick to fitting the global scale because
precisely estimating the drift electric field involves at least the clock
stripe sizes and their potentials during image integration. In this specific
study, we have to fit the overall model scale because we were not able
to find the drift and parallel clock potentials applied to the sensors.

\begin{figure}[ht]
  \begin{center}
    
\includegraphics[width=0.8\linewidth]{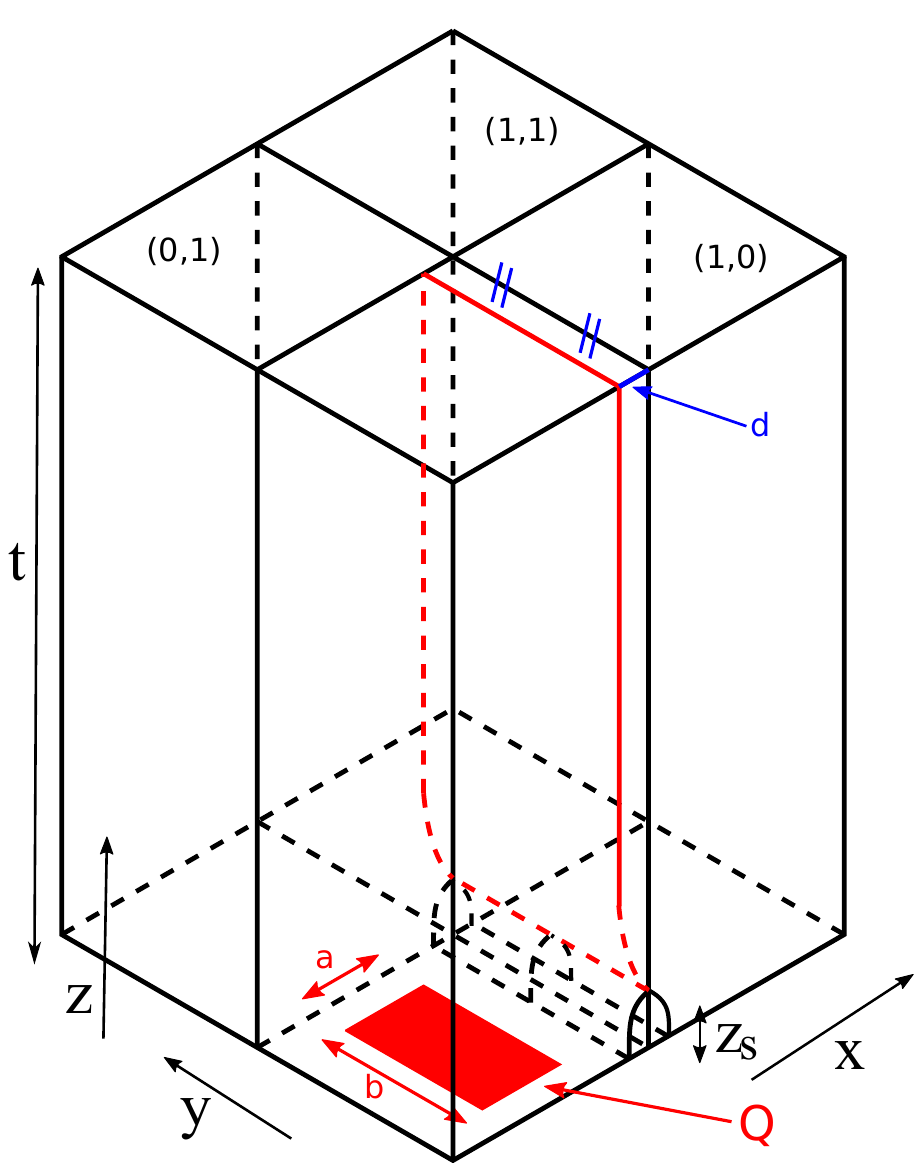}
\caption{Schematic of the geometry of the electrostatic fit.
     \label{fig:fig-drift2}
}
\end{center}
\end{figure}

We add some flexibility to the model by allowing source charges
to be extended and we stick to a very crude model: a uniformly
charged rectangle. This refinement, as compared to point sources,
only influences the very first neighbors. So the model eventually
has six parameters (see Fig. \ref{fig:fig-drift2}):
a global normalization factor, the height of the source charge $z_Q$, 
the height of the zero-field points over parallel and serial 
pixel boundaries $z_p$ and $z_s$, and the sides of the (uniform)
rectangular source charge $a$ and $b$.

The calculations of the model also require the thickness of the sensor
(200 $\mu$m) and the pixel side (15 $\mu$m). For the practical
implementation of the integrals from Eq. \ref{eq:model1}, we
settle for integrating analytically the terms of the series similar
to Eq. \ref{eq:potential} for the electric field and emulating
the rectangular source by splitting the charge into nine equally spaced
point charges (only for $n=-1,0,1$). We sum the series of the expressions in Eq.
\ref{eq:potential} up to $n=\pm 11$. This seems to be sufficient for
up to 10 pixels because increasing to $n=12$ only changes outputs at the sixth
decimal place. This electrostatic modeling approach was initially
presented in \cite{these-lebreton}, which we adapted
Figs. \ref{fig:fig-drift1} and \ref{fig:fig-drift2} from.

\begin{figure}[ht]
  \begin{center}
        
\includegraphics[width=\linewidth]{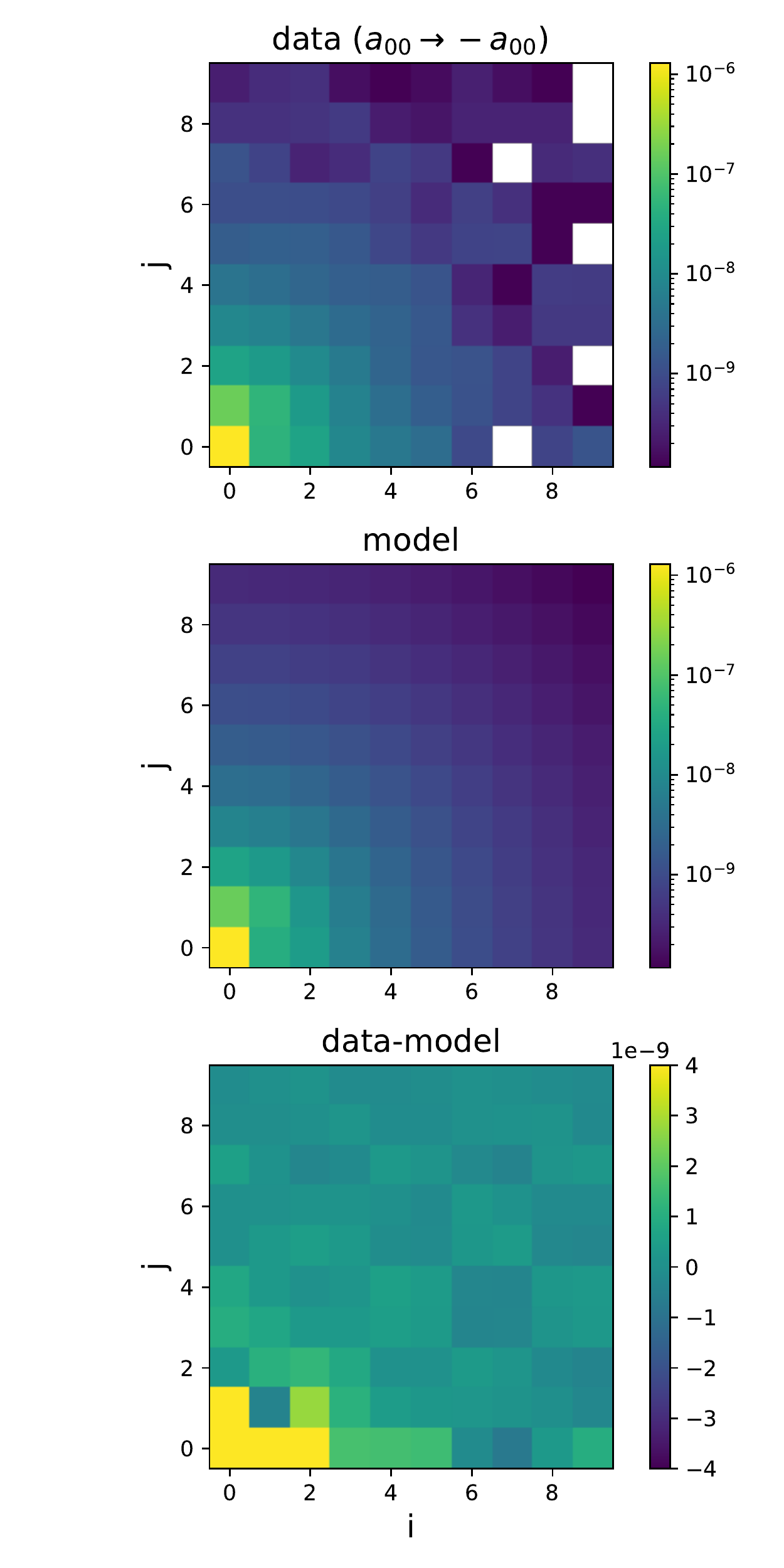}
\caption{Data (top), fitted electrostatic model (middle), and
  difference (bottom). The sign of $a_{00}$ has been flipped in the
  top plots, but not in the difference. We can note the clear
  difference of the two nearest neighbors, an anisotropy that is already
  visible in Fig. \ref{fig:ptc_and_co}. We expect a poor fit
  because the data violates the sum rule, while the fit cannot.
    \label{fig:elec-fit-model1}
}
\end{center}
\end{figure}

\subsection{Fit of the electrostatic model to the data}
\label{sec:electro_fit}
We performed a least-squares fit to the average area coefficient data,
using the spread over channels to weight the squares. We first fit the
average data obtained at the previous section. The data, fit, and
difference are shown in Fig. \ref{fig:elec-fit-model1}. We can see
that the measurements reach a signal-to-noise ratio of about 1 at a distance of 8
to 9 pixels from the source. Beyond the small separations, the
measurements exhibit a radial symmetry as expected from electrostatics
and displayed as well by the model. The residuals do not average to
zero and the data is larger than the model. This reflects that the
data violates the sum rule but the fit cannot. The sum of the model
area coefficients is $-7\ 10^{-8}$ up to $i,j<10$ (and tends to 0 with
increasing bounds, so the contribution of unmeasured $a_{ij}$ is
$7\ 10^{-8}$). Since the sum of measured $a_{ij}$ is $7\ 10^{-8}$, the
data violates the sum rule by about $1.4\ 10^{-7}$, which is more
than 10\% of $a_{00}$. The details of the electrostatic model cannot
change  the contribution of unmeasured $a_{ij}$ significantly (at $i>9$
or $j>9$) and certainly cannot flip its sign. The excess of the data
with respect to the model is concentrated on the 3 first serial pixels
and the first parallel neighbor.

\begin{figure}[th]
  \begin{center}
        
\includegraphics[width=\linewidth]{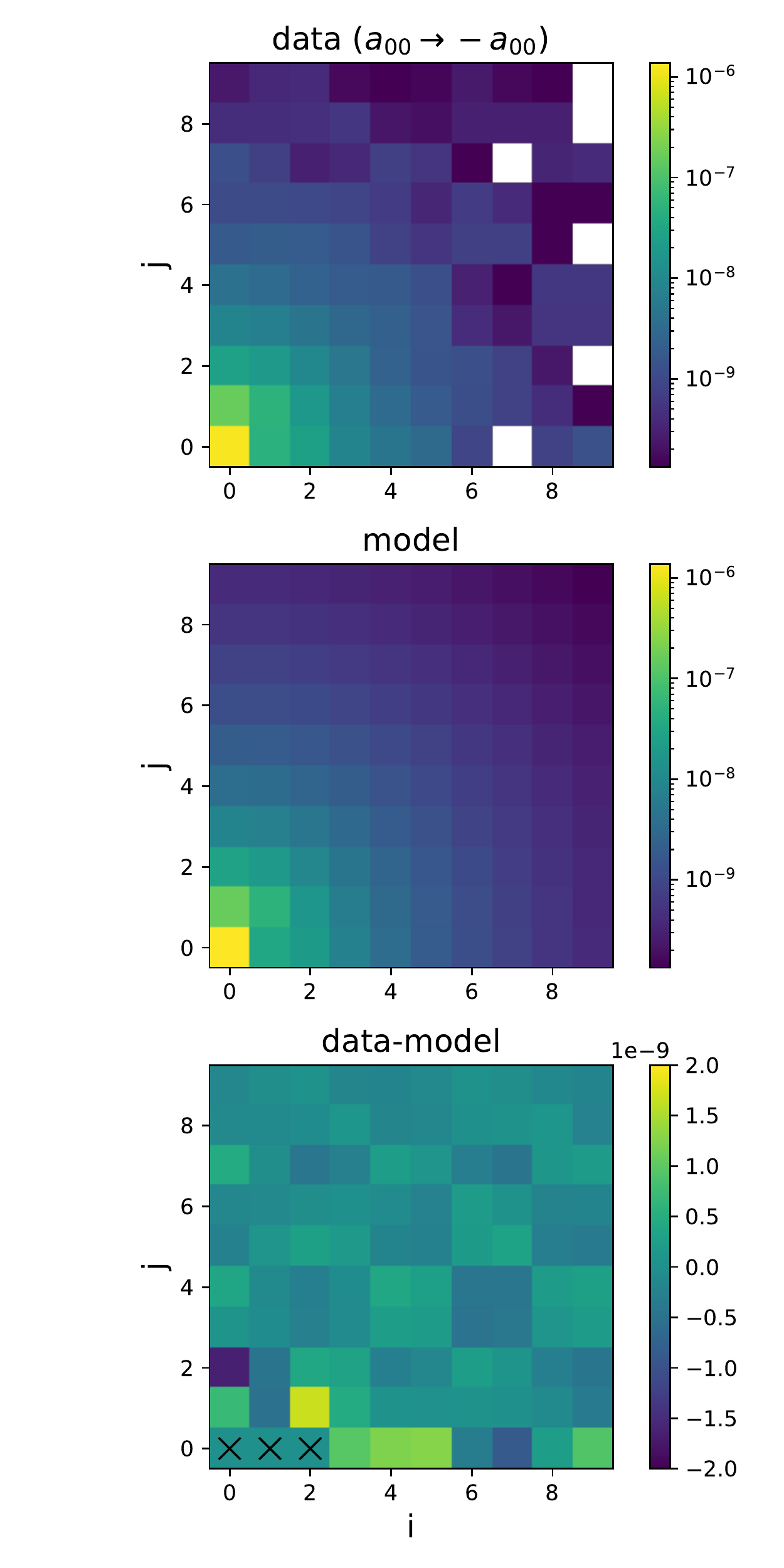}
\caption{Data (top), fitted electrostatic model (middle), and
  difference (bottom). This fit ignores the three first serial pixels,
  marked with crosses in the bottom plot. 
  Compared to Fig.~\ref{fig:elec-fit-model1}, we see that the data excess
  in $a_{01}$ has disappeared. We note that the color scale of the
  difference is zoomed-in, as compared to Fig.~\ref{fig:elec-fit-model1}
  \label{fig:elec-fit-model2}
}
\end{center}
\end{figure}
Some excess of variance and covariance along the serial direction is
expected if a video signal experiences rapid gain variation (or ``gain
noise''): rapid gain changes contribute a variance component that
scales as the square of signal level and, hence, artificially increase the value of
$a_{00}$ (see Eq. \ref{eq:C00}). If gain fluctuations last longer than the time to read a
pixel, they also bias covariances along the serial direction.
Since the residuals seem to decay along the serial direction,
possible gain variations have also to decay rapidly. They cannot
be invoked to explain the excess on the first parallel neighbor because
while serial pixels are read out microseconds apart, milliseconds
separate neighboring lines.

We now attempt a fit that ignores the variance and the two next serial
pixels, that is, $a_{00}$, $a_{10}$ and $a_{20}$.  The fit displayed in
Fig.~\ref{fig:elec-fit-model2} seems acceptable: the residuals are at
most 1\% of the largest used coefficient $a_{01} \simeq 1.6\ 10^{-7}$,
and an even lower fraction of $a_{00}$. The residuals exhibit some sort
of low-level ``chessboard pattern'', which is not entirely systematic.
Assuming it is real, we have no proposition for its source, and we 
could not invent a small periodic variation of the physical size of pixels
that would produce this $2\times 2$ pixel pattern of the two-pixel
correlation functions. 

\begin{figure}[ht]
  \begin{center}
        
\includegraphics[width=\linewidth]{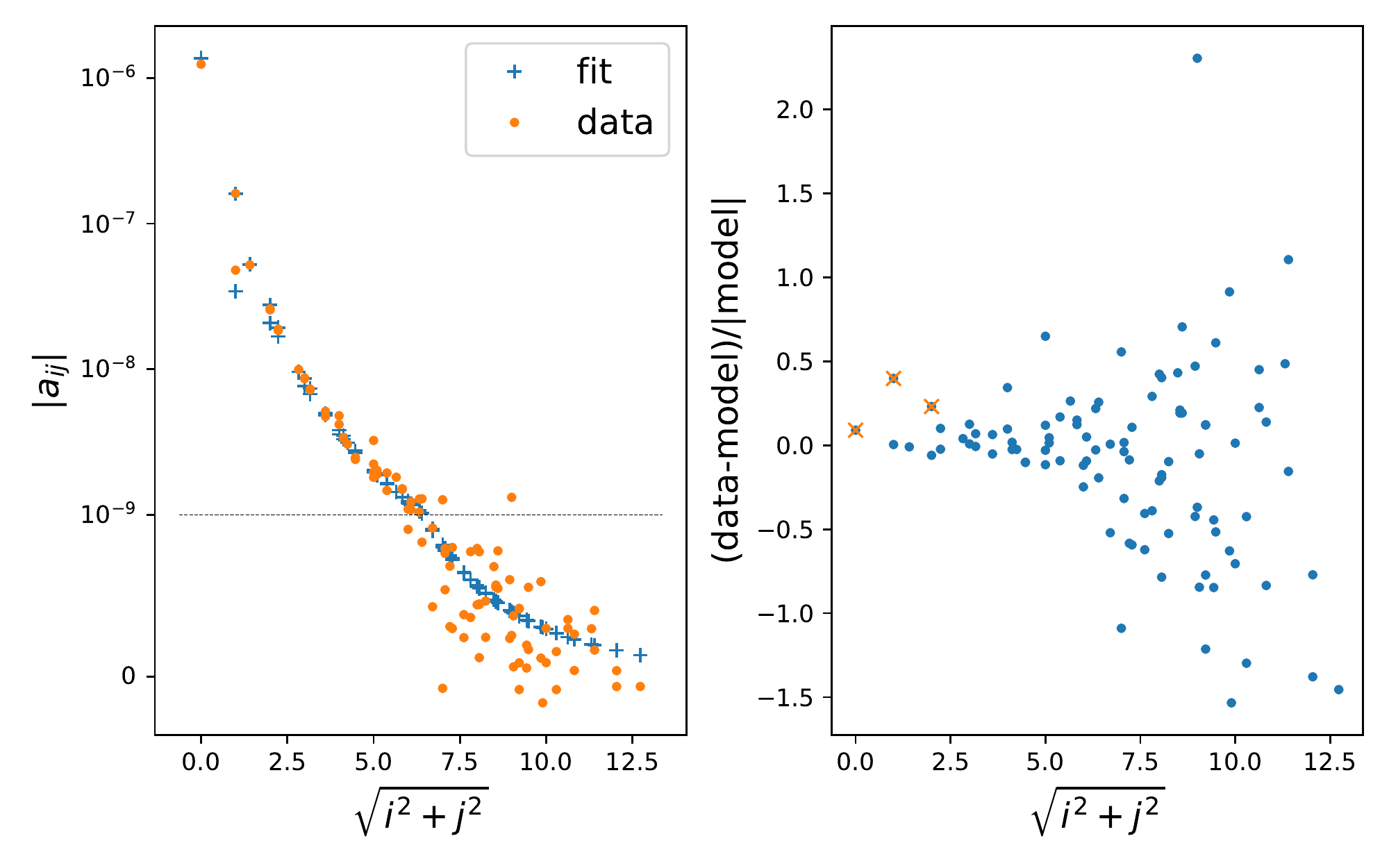}
\caption{Data and fit (left) as a function of distance, difference
  normalized to the model (right). The fit ignores the three first
  serial pixels, which are labeled with crosses in the rhs plot. The dashed line
  in the lhs plot separates the logarithmic and linear scale regions.     
  \label{fig:radial_noserial}
}
\end{center}
\end{figure}

We display the measured and fitted values, and their ratio in Fig.~
\ref{fig:radial_noserial}. The model reproduces the data well and
gives some confidence that the model delivers sensible values for the
data we decided not to use. We note that the decay of signal with
distance is reproduced very well  by the model over more than two
orders of magnitude of signal level. In \cite{Pumplin69}, it is shown
that the large-distance decay of the perturbating electric field from
a point charge in the sensor depends essentially exponentially on the
inverse thickness of the sensor. The other geometrical parameters of the model
cannot alter significantly the logarithmic slope of this decay at
large distances. So, the evaluation of the model at large distances, needed
to gauge the compliance of data to the sum rule, is a robust
outcome of the model if the slopes of data and model match. 

\begin{table}[h]
  
\caption{Values of the largest area coefficients for the data and two fitted models}
\begin{center}
\begin{tabular}{|l|l||r|r|r|}
    \hline    
    & &  \multicolumn{1}{c|}{$i=0$} & \multicolumn{1}{c|}{$i=1$} & \multicolumn{1}{c|}{$i=2$} \\
    \hline
    \hline
      & d  & $2.60 \ 10^{-8}$ & $1.88 \ 10^{-8}$ & $9.95 \ 10^{-9}$ \\
$j=2$ & m1 & $2.57 \ 10^{-8}$ & $1.78 \ 10^{-8}$ & $8.66 \ 10^{-9}$ \\
      & m2 & $2.77 \ 10^{-8}$ & $1.93 \ 10^{-8}$ & $9.57 \ 10^{-9}$ \\
\hline
      & d  & $1.61 \ 10^{-7}$ & $5.19 \ 10^{-8}$ & $1.85 \ 10^{-8}$ \\
$j=1$ & m1 & $1.55 \ 10^{-7}$ & $5.23 \ 10^{-8}$ & $1.57 \ 10^{-8}$ \\
      & m2 & $1.60 \ 10^{-7}$ & $5.24 \ 10^{-8}$ & $1.68 \ 10^{-8}$ \\
\hline
      & d  & $-1.24 \ 10^{-6}$ & $4.79 \ 10^{-8}$ & $2.56 \ 10^{-8}$ \\
$j=0$ & m1 & $-1.29 \ 10^{-6}$ & $3.90 \ 10^{-8}$ & $1.99 \ 10^{-8}$ \\
      & m2 & $-1.37 \ 10^{-6}$ & $3.43 \ 10^{-8}$ & $2.08 \ 10^{-8}$ \\
\hline
\end{tabular}
\end{center}
\tablefoot{The three rows for each value of $j$ refer to the data, model 1, and model 2. Model 1 fits all the measured $a_{ij}$, while model 2 ignores the data from the bottom row. \label{tab:area_coeff}}
\end{table}

In Table \ref{tab:area_coeff}, we display the measured and fitted
largest area coefficients of the data and the two fitted models. We
can see in particular the size of the discrepancies along the serial direction.
We can now extract from the fitted model the pixel boundary
displacement caused by a unit charge. This allows us to compute for a
given science image, the accumulated boundary displacements resulting
from all present charges. We will test the quality of the BF effect
correction derived from the models on science images in the following section.
We now question the curl-free hypothesis, and then discuss the gain noise
issue.

  \subsection{Considering whether the pixel boundary displacement field is curl-free}
  \label{app:curl_free}

  \begin{figure}[ht]
  \begin{center}
        
    \includegraphics[width=\linewidth]{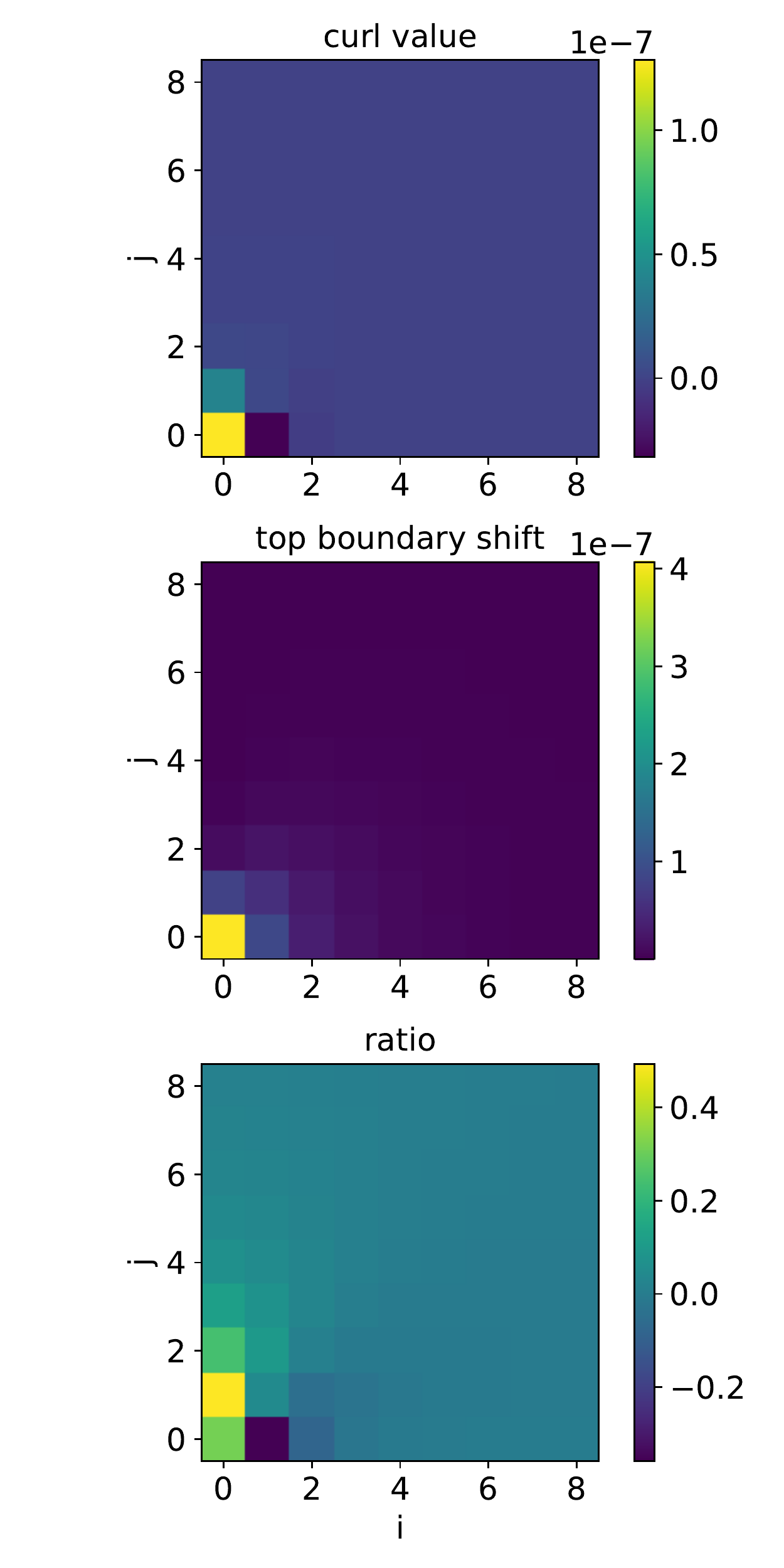}
    \caption{Values of the (discrete) curl, the top boundary displacement
      and their ratio. One can see that assuming a curl-free displacement
      field is acceptable only at distances larger than $\sim$ 3 pixels. 
      \label{fig:curl}
    }
    
  \end{center}
  \end{figure}
    In C18, the authors assume that the boundary displacement field is
    curl-free. Our definition of discrete curl is expressed as:
    \begin{equation}
      c_{i,j} = (a^N_{i,j+1} - a^N_{i,j}) - (a^W_{i+1,j} - a^W_{i,j}).
    \end{equation}
    In Fig. \ref{fig:curl}, we display the discrete 2D curl
    of the field, the displacement of one boundary (arbitrarily chosen
    as the parallel one, the serial one is in fact smaller) and their
    ratio. We used the electrostatic solution displayed in
    Fig. \ref{fig:elec-fit-model2}, which (by definition) satisfies the
    Poisson equation and, hence, has a curl-free 3D electric
    field. We can readily see that the curl-free assumption may lead
    to sizable errors (several tens of percent) in the correction.

    One reason for the displacement field to be rotational is that
    drift paths along serial and parallel boundaries do not end at the
    same distance from the parallel clock stripes. The importance of this
    difference vanishes with distance because at large distances, the
    field varies less steeply with $z$ than at short distances and,
    hence, the fractional contribution to the boundary displacement of
    the end of the drift path decays with distance. When it comes to
    science images, and especially when the image quality is good, the
    short-distance boundary displacements dominate the effective
    correction for stars.

\subsection{Gain noise}
We measure $a_{00}=-1.24 \ 10^{-6}$, while the fit of
Fig. \ref{fig:elec-fit-model2} indicates $a_{00}=-1.37 \ 10^{-6}$. This
means that the model predicts a variance that is smaller than the
measured one, and the leading difference is quadratic in signal
level. This difference of $a_{00}$ values also contributes to the 
apparent curvature of the residuals displayed in
Fig.~\ref{fig:plot_res}: it roughly explains the curvature for $C_{00}$ and $C_{01}$ -- but not for $C_{10}$.

While questioning the area coefficients along the serial direction,
it is tempting to relate the excess we observe to charge
transfer issues; however, we provide a few arguments against this
explanation. First, we have measured and corrected deferred signals
and we find that the linear dependence displayed in Fig. \ref{fig:cti}, causes, if
uncorrected, a linear contribution to $C_{10}$ ($\propto \mu$); however, the
$a_{10}$ excess we observe corresponds to a quadratic contribution to
$C_{10}$ ($\propto \mu^2$). Second, any mechanism relying on imperfect
charge transfer will unavoidably affect $C_{00}$ and $C_{10}$ in
comparable and opposite amounts, but the data in Table \ref{tab:area_coeff}
indicate that the excess of $a_{00}$ is about 20 times larger that
the one of $a_{10}$, and they have the same sign. Third, serial charge
transfer inefficiency is usually associated to localized and rare
defects that is typically 1 or more frequently 0 along the serial
register.  The distributions of area coefficients (Fig.
\ref{fig:plot_a_dist}) do not seem quantized.

So, assuming that this difference of $a_{00}$ values is due to gain variations
during the read out, the relative gain variations should be about
$3.5\ 10^{-4}$ rms. Gain variations of the HSC electronics have been
studied during the HSC camera fabrication process and are reported in
\cite{Miyakate12}. They are evaluated as $\sim 2.4 \ 10^{-5}$, however,
this is only considering the video chain and not the CCD (and perhaps
in a context that is different from what the instrument actually faces).

The CCD readout chain of HSC relies on correlated double sampling, an
approach that integrates the video signal during logical gates
provided by the clocking system. The collected signal is hence
vulnerable to fluctuations of the timing of the integration gates. We
cross-correlated the sub-images from different channels of the same sensor
in order to diagnose synchronous correlations, including variations of
the gains (whatever the cause). We did not find any compelling
signal for any sensor. We were hoping that cross-correlations
would deliver the size of gain fluctuations, or any other common-mode
fluctuation of channel response, thus allowing us to subtract
this contributions from flat-field statistics.

In summary, we are not able to provide a convincing cause of the sum
rule violation. Because the electrostatic fit indicates an excess of
fluctuations along the serial direction, we ignore three area
coefficients. This excess is obviously puzzling, albeit compatible with rapid
gain fluctuations (with some ad hoc time decay). Fortunately, we do
not have to rely in any way on this hypothesis to use the
electrostatic model.

\section{Correction of science images}
\label{sec:sci_corr}
\subsection{Data set and reduction}

We process the images of the Cosmos field from the ultra-deep part of the Subaru
Strategic Program,  acquired between 2014 and
2019.
The ultra-deep part of the survey is geared
at obtaining extremely
deep images by co-adding a large number of observations, as well as
detecting high-redshift supernovae and measuring their light curves. The
sample of images covers the five bands: $g$, $r$, $i$, $z$ and $y$ of the
camera, and also covers a broad range of image qualities. We use data
acquired with the new $i$ and $r$ filters, which are called $r2$ and
$i2$ in HSC parlance and we stick to these names. This image sample
constitutes a representative playground for testing the quality of the
brighter-fatter correction on HSC science images.

The image reduction we perform is fairly standard for the brighter-fatter
correction. We first average the overscan and subtract it from the
actual image data. We then correct each image segment for non-linearity.
Then we have to express the image in electrons in order to
perform the BF correction, so we multiply each image segment by its
corresponding gain (as determined when fitting the covariance data),
perform the correction, and divide back each segment by the gain. We note that the BF
correction has to deliver an image that has not been corrected by channel
gains because our flats contain the gain differences and, hence, they should
be applied to an unscaled image. We prefer to use flats that encode
the relative gains
because their evolution with time encodes a possible evolution of gains.
Using approximate gains for the BF correction is clearly a second-order
issue, while using approximate relative gains on a sensor can cause
artificial steps in the sky background at the channel boundaries.

The brighter-fatter correction consists in computing the boundary shifts
by summing all the actions of all image charges on all boundaries.
For parallel boundaries, the shifts are expressed as:
\begin{equation}
  \delta_{ij}^N = 1/2\sum_{kl} a^N_{k,l} Q_{i-k, j-l}
  \label{eq:boundary_shifts}
,\end{equation}
and similarly for serial boundaries. The factor  of $1/2$ accounts for the
fact that source charges alter pixel shapes only once they have
reached the pixel wells; thus, on average, during half of the integration
time. In Eq. \ref{eq:delta_a}, the charge on the rhs refers to
the charge accumulated so far during the integration and, hence, the time-averaged
boundary shift should be derived from the time-averaged charge content.
The charge to displace from one pixel to its neighbor is
computed from the pixel boundary shift and the charge flowing over the
same pixel boundary. We compute the latter as the average between the
two pixels that share the boundary.  We tried a quadratic order
interpolator (involving 4 pixels), and did not find a decisive
difference. We should note that the vast majority of our images are
well sampled (the image quality is larger than 3 pixels) and sharper
images could react differently. Ultimately, the correction of the BF
effect in the images themselves requires that an accurate estimation
of the charge at pixel boundaries can be devised.

Once the BF correction has been applied, we divide back each channel
section by its gain and apply the flat field constructed from dome
flats accumulated over about one month. For the $y$ band, we have
to subtract a fringe pattern constructed from a large set of science
images.

On each individual CCD from each exposure, we
run SExtractor \citep{Sex} to detect objects and measure the
Gaussian second moments of these objects from an unweighted 2D
Gaussian fit to the light distribution. Because we do not integrate
the Gaussian over pixels, we use the fast procedure described in
\cite{Astier-photom13} to solve the normal equations. These second
moments are similar if not identical to the ``SDSS adaptive moments.''
We identify stars in the image from the distribution of
moments (see \citealt{Astier-photom13}). Because we have
to accommodate significant variations of the PSF over the field
of a single CCD, the cut that selects stars is broad enough
to avoid rejecting genuine stars because of the BF effect.
We get three moments per star: $M_{XX}$, $M_{YY}$ and $M_{XY}$,
where $X$ and $Y$ refer to the serial and parallel directions
on the sensor. As an indicator of PSF size, we use:
\begin{equation}
  IQ_2 \equiv T_{PSF}/2 = (M_{XX}+M_{YY})/2.
\end{equation}

We then need color measurements of our stars, which we obtain by
averaging fluxes over images and calibrating instrumental magnitudes
over the common footprint of our star catalog and the field D2 from
\cite{Betoule13}. This is not a
critical step since only small color corrections are involved in what
follows and we, hence, do not need colors measured to better than
$\sim$0.1~mag.

\subsection{Processings and results}
\begin{figure}[ht]
  \begin{center}
        
\includegraphics[width=\linewidth]{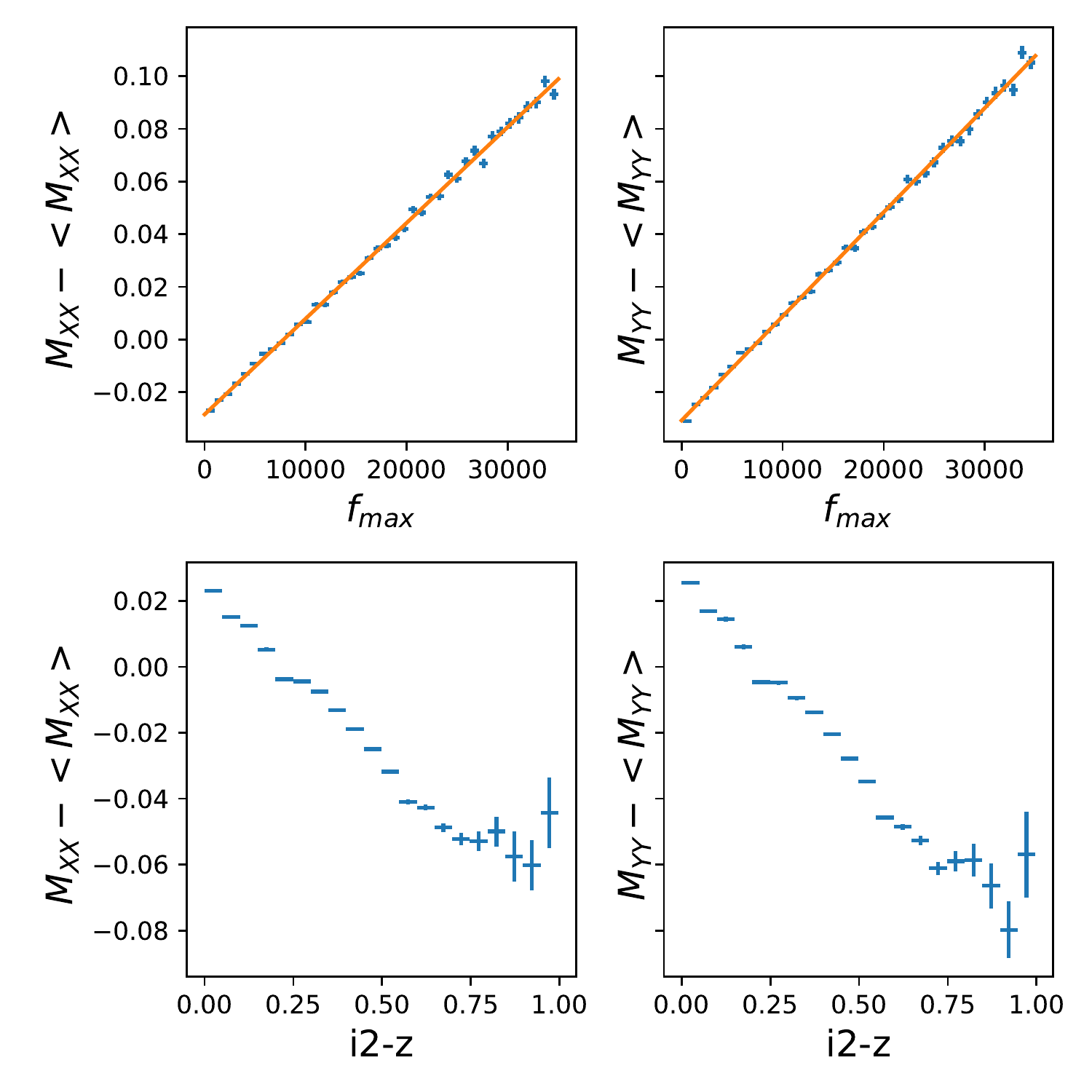}
\caption{Differences between the star second moments and their
  expectations from the spatial smoothing, as a function of $f_{max}$
  (top) and color (bottom). We have selected stars with $3<IQ_2
  <4\ \mathrm{pix}^2$, in the $z$ band. The color range was chosen so
  that the color dependence is roughly linear. One can see that the
  slope of the top right plot is slightly steeper than the top left
  plot, reflecting the anisotropy of pixel covariances in flat fields.
  \label{fig:actual_bf}
}
\end{center}
\end{figure}
In order to assess the variation of star moments with their flux, we
report the variation of star sizes with peak flux $f_{max}$, rather
than with the flux itself. The variation with $f_{max}$ is of practical
interest, because $f_{max}$ determines if a given star can enter into
the PSF modeling. In order to isolate the variation of moments with
peak flux we have to account for two other variations: the spatial
variation in every exposure (mostly due to optics), and the color dependence of apparent star
sizes.  We remove the spatial dependence by fitting a second-order 2D
polynomial to the star moments measured on a given CCD in a given
exposure. We then interpolate this crude model at the star position
and study the residual of the measurement to the model. We model
 $M_{XX}$ and $M_{YY}$independently. We apply a conservative cut at
$f_{max}<35000$~ADU, in order to avoid any effect of sensor
saturation. The dependence on $f_{max}$ and color of these residuals
are shown in Fig. \ref{fig:actual_bf}, for a processing in which no BF
correction was applied. We see that for stars selected at
($M_{XX}+M_{YY})/2 \simeq 3.5\ \mathrm{pix}^2$,  the total increase of
moments with $f_{max}$ is $\sim 0.13\ \mathrm{pix}^2$, which is about
3.7\% for stars varying from 0 to saturation. This slope depends on the
selection of image quality. Figure \ref{fig:actual_bf} also indicates that the
variation of apparent size with color in the $z$ band are not
considerably smaller than the ones with flux. Since the
$f_{max}$-color correlation coefficient is about $-0.18$ (in $z$ band),
we should account for color-induced size variations.

In order to account for both the peak flux and color dependence of
the moments, we regress the star moment residuals against both $f_{max}$ and color:
\begin{equation}
  M_{AA} - M_{AA}^{expected} = \alpha f_{max} + \beta c + \gamma
  \label{eq:double_fit}
,\end{equation}
where $A$ stands for $X$ or $Y$. We then use $\alpha$ (i.e., the BF
slope) as a color-independent indicator of the BF effect intensity
(both before and after correction). The linear correction in color can only
work within a finite color range, and the selected color indicators and
their range are provided in Table \ref{tab:color_ranges}. For this
fit, we ignore the measurements with $f_{max}$<5000 because the
measured moments could be affected by a biased background estimation.
Out of about $15\ 10^6$ star measurements, the flux and color cuts
retain $2.4\ 10^6$ measurements contributing to the slope measurements that are reported here. 
\begin{table}[h]
  
\caption{Color indicators and color ranges used for the various HSC bands}
\begin{center}
\begin{tabular}{|l|l|c|c|}
  \hline
band & color & $c_{min}$ & $c_{max}$ \\
\hline
g & g-r2 & 0.3 & 1.1 \\
r2 & r2-i2 & 1.0 & 2.0 \\
i2 & i2-z & 0.0 & 1.0 \\
z & i2-z & 0.0 & 1.0 \\
y & z-y & 0.0 & 0.5 \\
\hline
\end{tabular}
\end{center}
\tablefoot{For each band, the analysis selects stars with color
  between $c_{min}$ and $c_{max}$ in order to perform the regression
  of Eq. \ref{eq:double_fit}.
\label{tab:color_ranges}}
\end{table}

In order to study the quality of the BF correction on real science
conditions, we bin stars in moments bins and restrict
the range to $(M_{XX}+M_{YY})/2<10$, which roughly corresponds to an
image quality of 1.35\arcsec~FWHM. We do not apply a lower
cut because the best image qualities are precious for at least
cosmic shear. The bin selection operates on the expected moments at the
location of the star rather than the measured ones, so that the
quantity used to select the bins contents is statistically independent of flux
and color. 
\begin{figure}[ht]
  \begin{center}
        
\includegraphics[width=\linewidth]{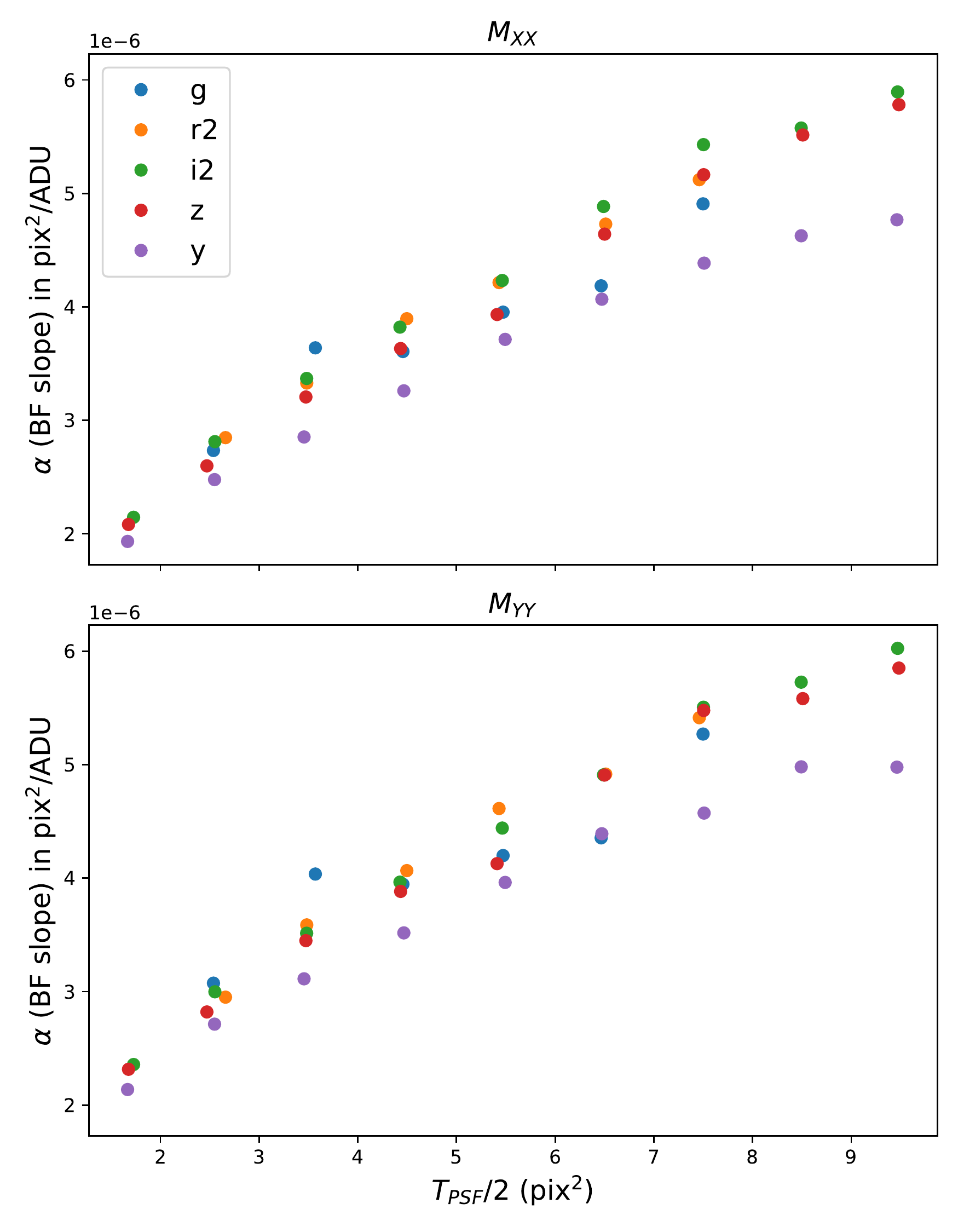}
\caption{BF slopes (with color correction) in $IQ_2$ bins for the five
  bands of HSC, without any BF correction, for the serial (top) and
  parallel (bottom) directions. We can note a roughly linear increase
  of the slopes with $IQ_2$, and that for the best image qualities,
  the slope for $M_{YY}$ is larger than for $M_{XX}$, as expected from
  the anisotropy of the nearest neighbor $a_{ij}$ (as seen in
  Fig. \ref{fig:elec-fit-model1})
  \label{fig:bf_slopes_no_bf}
}
\end{center}
\end{figure}

\begin{figure}[ht]
  \begin{center}
        
    \includegraphics[width=\linewidth]{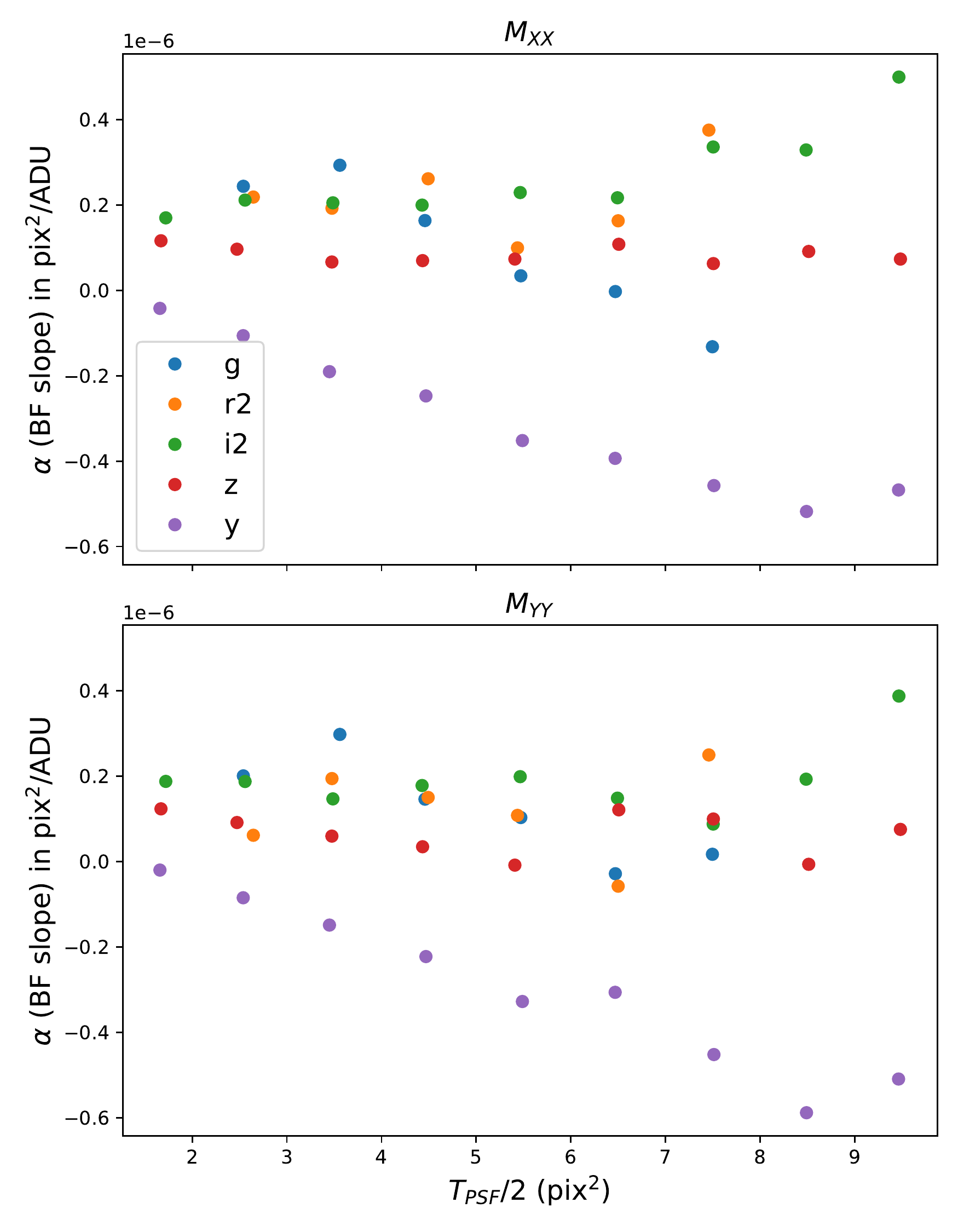}
\caption{BF slopes (with color correction) in $IQ_2$ bins for the five  bands of HSC, with the BF correction derived from
  ``model 1'' (Fig. \ref{fig:elec-fit-model1}) which uses all
  the covariance measurements. One can note a residual BF slope
  essentially independent of image quality, for all bands, except $y$
  that deserves a specific treatment (detailed in \ref{subsec:y-band-model}).
  \label{fig:bf_slopes_model1}
}
\end{center}
\end{figure}

The measurements of $\alpha$ (accounting for color corrections) for
both CCD directions and in $IQ_2$ bins are displayed in
Fig. \ref{fig:bf_slopes_no_bf}.  We can notice an increase of the
slopes with $IQ_2$, as well as the fact that the $y$ band is notably less affected by
the BF effect than the other bands. In the $y$ band, the photons convert deeper
in the sensor bulk and the charges experience a shorter drift
path than for bluer bands, thus reducing the action of the
perturbating electric field. For the $y$ band, the starting point of
the integral of Eq. \ref{eq:model1} is noticeably smaller that the
sensor thickness $t$.

\begin{figure}[h]
  \begin{center}
        
    \includegraphics[width=\linewidth]{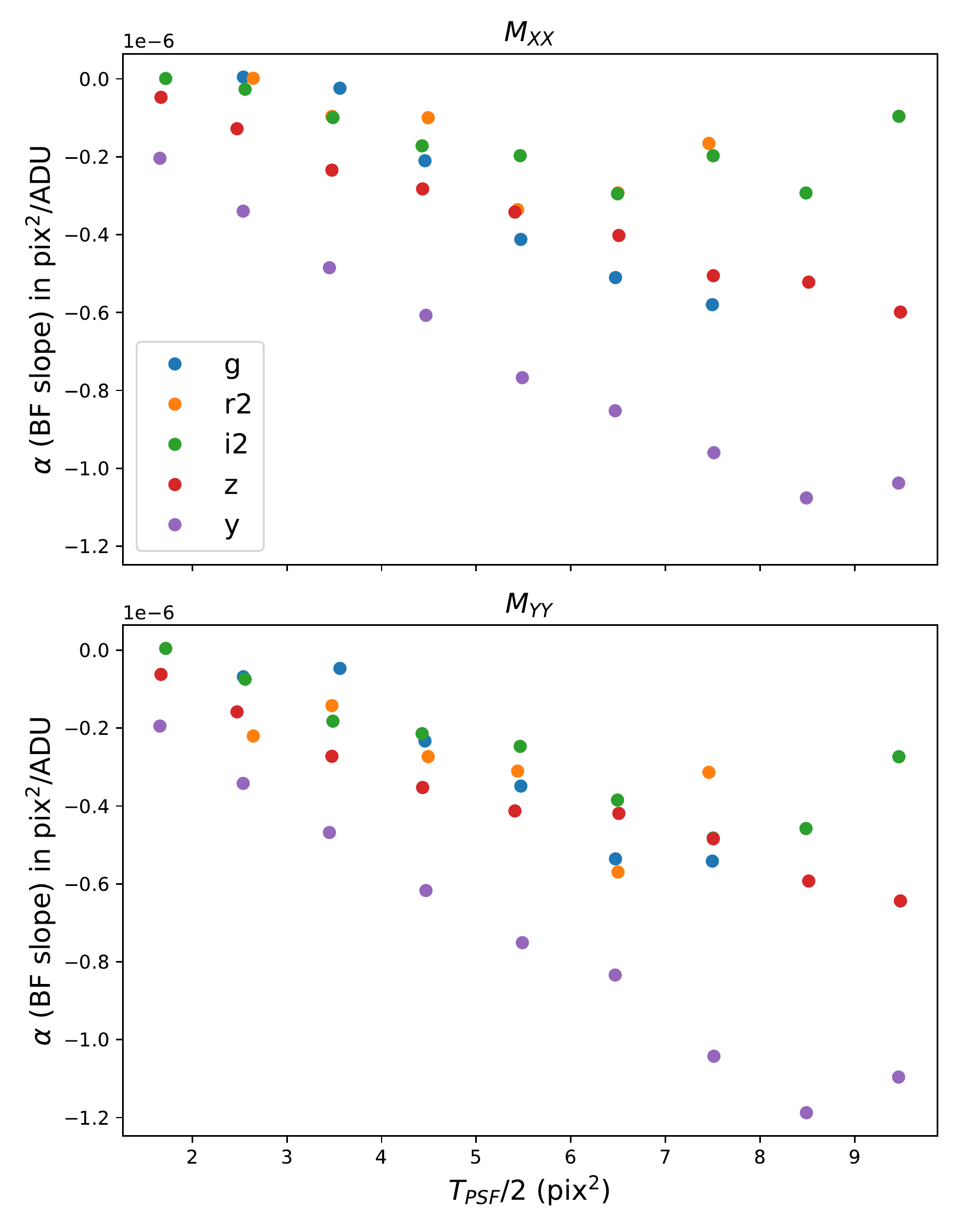}
\caption{BF slopes (with color correction) in $IQ_2$ bins for the five  bands of HSC, with the BF correction derived from ``model 2''
  (Fig. \ref{fig:elec-fit-model2}) which uses all the covariance
  measurements but the three first serial pixels.  We face a small
  over correction, and the BF slopes have been reduced by a factor of
  about 30 as compared to the raw values of
  Fig.~\ref{fig:bf_slopes_no_bf}. The $y$ band still requires a specific
  treatment. 
  \label{fig:bf_slopes_model2}
}
\end{center}
\end{figure}

For some time, we ignore this subtlety and correct all bands using a
same model for all bands, derived from covariances measured in the $g$ band.  We first apply the correction derived from ``model 1'' (displayed
in Fig. \ref{fig:elec-fit-model1}) and the residual BF slopes are
displayed in Fig. \ref{fig:bf_slopes_model1}.  Although the BF slopes
are significantly reduced with respect to the raw data
(Fig. \ref{fig:bf_slopes_no_bf}), the residual slopes are large at
low $IQ_2$, indicating that the correction is underestimated at small
distances. This correction leaves about 10\% of the BF effect at the best
image qualities and about 3\% at the upper end. 

We then apply the ``model 2'' correction (Fig.~
\ref{fig:elec-fit-model2}), namely, the one that ignores the three first
suspicious measurements along the serial direction. The corresponding
BF slopes are displayed in Fig.~\ref{fig:bf_slopes_model2}. The
quality of the correction, in particular at the lowest IQ is
improved. Ignoring the $y$ band, we can interpret the figure as a
global small overcorrection that leaves BF slopes about 30 times
smaller than in the raw data.

\subsection{Brighter-fatter correction for the y-band}
\label{subsec:y-band-model}
When the energy of photons becomes comparable to the silicon band gap,
the absorption cross-section tends to vanish and silicon becomes
transparent. The band gap is about $E=1.2$~eV corresponding to
$\lambda=1.1$~$\mu m$. In the $y$ band, the absorption length of
photons becomes comparable to the sensor thickness and we can no
longer assume that charges produced by converted photons drift all the
way from the entrance window (z=$t$ in the coordinates of \S
\ref{sec:elec-model}). The flat-field data was acquired in $g$ band,
where the mean free path of photons is well below 1~$\mu m$, and
the approximation of immediate conversion is adequate. This
approximation works up to $i$ band, may be questioned for $z$ band
and does not seem to apply to $y$ band.

Once we have our electrostatic model, the prediction for boundary motions in $y$
band can be computed by altering the lower bound of the integral of Eq.
\ref{eq:model1} :
\begin{equation}
  d(z_c) = k \int_{z=z_c}^{z=z_0} E_Q^T(x_b,y_b,z) dz
  \label{eq:model_y}
,\end{equation}
where $z_c$ is where the photon converts and $k$ is the normalization
determined when fitting the model to the area coefficient data
(obtained in $g$ band in our case). Since the integrand can vary rapidly with $z$, $\overline{d(z_c)} \neq d(\overline{z_c})$, it is unwise to evaluate these integrals at the average conversion depth of photons in the $y$ band. We instead compute the
average of $d(z_c)$ for a realistic distribution of $z_c$:
\begin{equation}
  d_y = \int_{z_c=t}^{z_c=sz_b} d(z_c) dN/dz_c\ dz_c, 
\end{equation}
where $dN/d\ z_c$ is the distribution of conversion points (and should
integrate to unity over the integration domain), and $z_b$ refers to
$z_s$ or $z_p$, depending on which boundary we are considering. The
distribution of conversion depths depends on photon absorption length and some
assumption for the object spectrum. The latter may be regarded as
inconvenient, but this is a small chromatic dependence of a small
flux dependence and, hence, a second-order effect.
\begin{figure}[h]
  \begin{center}
        
    \includegraphics[width=\linewidth]{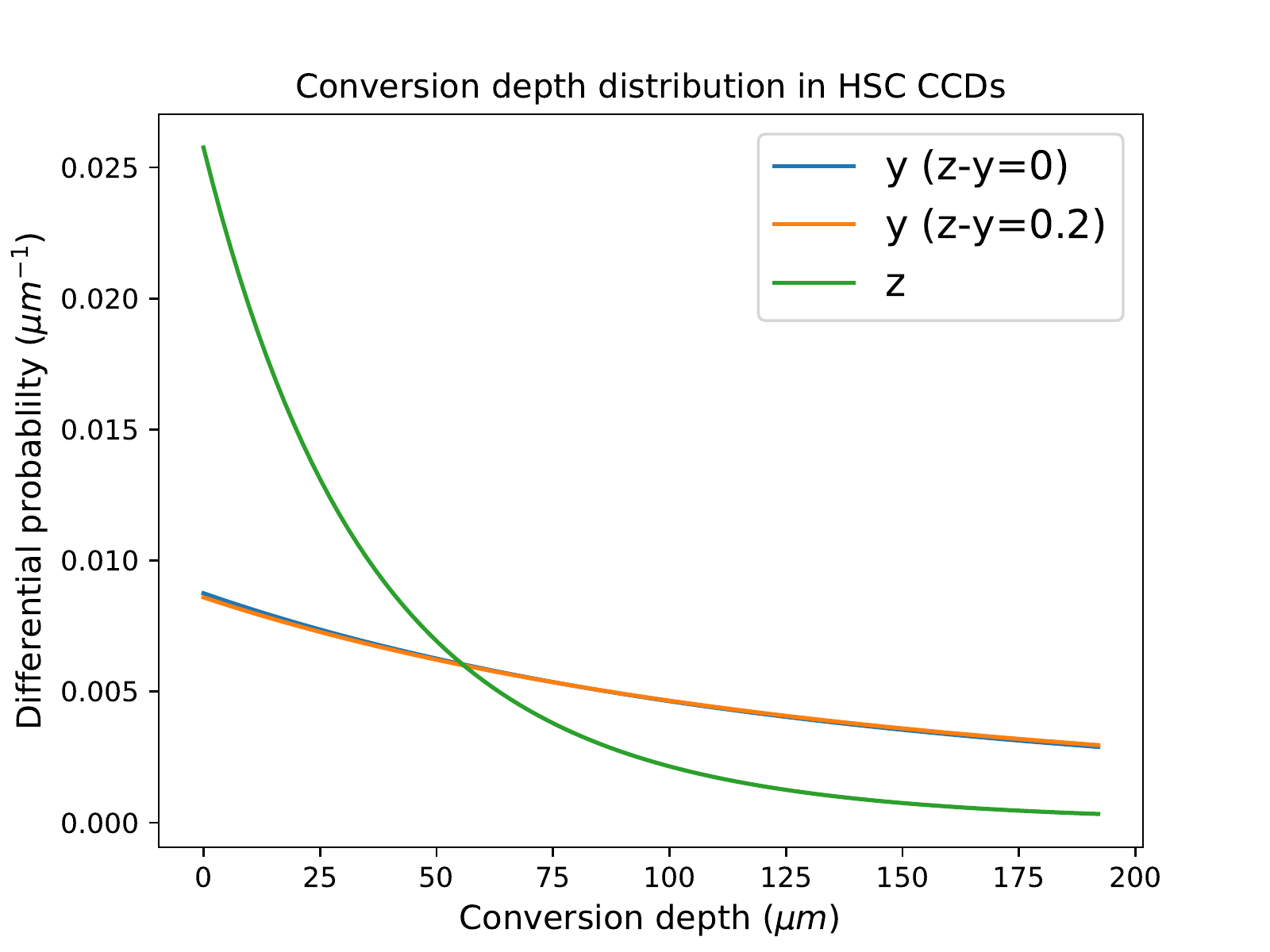}
    \caption{Distributions of conversion depth for a zero color object in
      the HSC CCDs. The $y$ distribution is much flatter and leads to a
      globally reduced BF effect. Changing the color of the object by
      about the width of the our $z-y$ distribution does not
      significantly alter the expected depth distribution in the $y$
      band. For the $z$ band, most of the conversions occur
      at small depths, where the perturbating electric field is
      small. For the $i2$ band, the average conversion depth is
      about 10~$\mu$m.
  \label{fig:conversion_depth}
}
\end{center}
\end{figure}

\begin{figure}[h]
  \begin{center}
        
    \includegraphics[width=\linewidth]{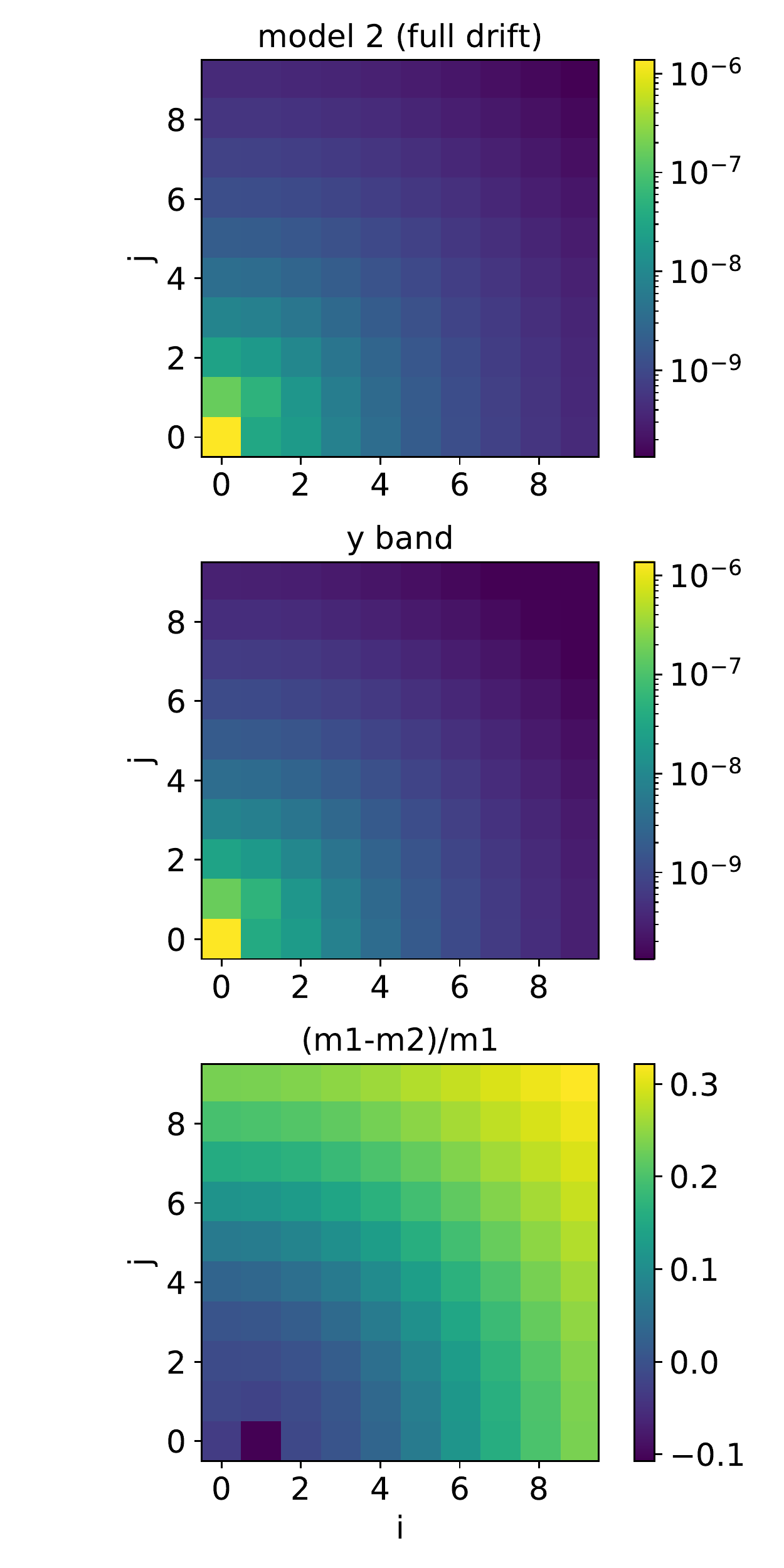}
      \caption{Area coefficients for ``model 2'' (displayed in
        Fig. \ref{fig:elec-fit-model2}) based on the same model with shorter drift
        paths meant to describe the $y$ band (center) and the
        relative difference at the bottom. The differences tend to increase with
        distance to the source.
  \label{fig:model_y_diff}
}
\end{center}
\end{figure}

We choose the spectrum of a color 0 (in AB magnitudes) object to
compute the conversion depth distribution, noting that $z-y=0$ lies
inside the observed star color distribution, which has an average of
$<z-y>\simeq 0.13$. For the absorption length as a function of
wavelength, we use the expression from \cite{Rajkanan79} for Silicon
at $173 ^oK$, which typically predicts 103~$\mu m$ at
$\lambda=950$~nm. The conversion depth distributions for $z$ and $y$
bands is displayed in Fig. \ref{fig:conversion_depth}, and we can note
that in $y$ band, the conversions are spread over the whole
thickness and the depth distribution is stable when changing the
color by about the width of the $z-y$ distribution of stars. In
Fig. \ref{fig:model_y_diff}, we compare the electrostatic model 2 integrated over the
full thickness with the same model for the $y$ band, using Eq.
\ref{eq:model_y}, with the distribution of Fig.~\ref{fig:conversion_depth}. The largest relative changes happen at large
distances, where the perturbating electric field has a sizable
contribution over most of the drift path.

\begin{figure}[h]
  \begin{center}
        
    \includegraphics[width=\linewidth]{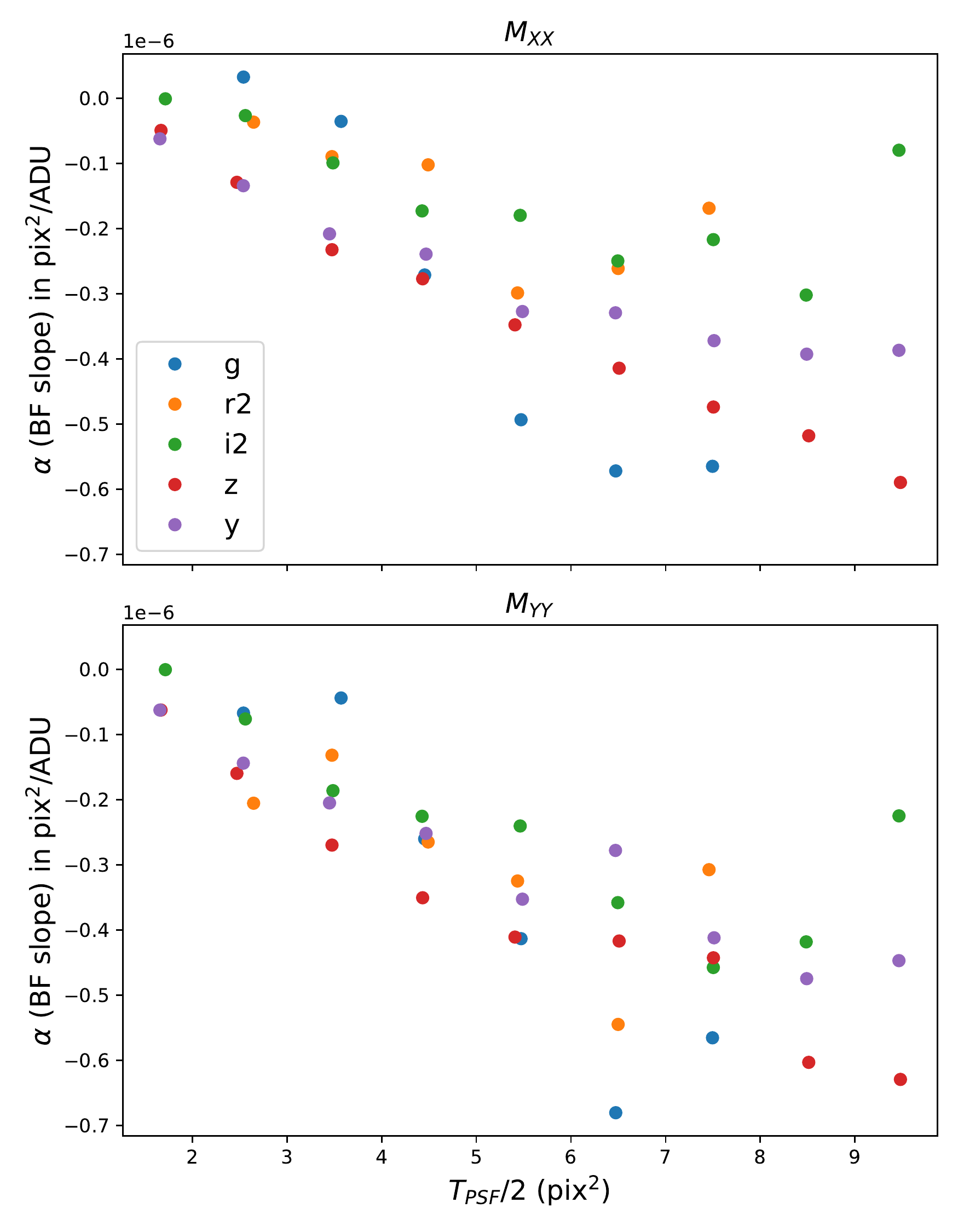}
    \caption{BF slopes computed after correcting images with model 2,
      with the modifications described above for $y$ band. With respect to
      Fig.~\ref{fig:bf_slopes_model2}, only the $y$ data has changed. We
      may note that all bands now behave in a similar way.
  \label{fig:bf_slopes_model2_y}
}
\end{center}
\end{figure}

Next, we apply this modified BF correction for the $y$ band to the
actual science data. We can see in Fig. \ref{fig:bf_slopes_model2_y} that the
$y$ band thus behaves similarly to the other bands, which indicates
that the reduced BF effect in $y$ likely originates in shorter drift
paths in this very red band. This also indicates that for these CCDs, flat field
correlations measured in a blue band can be used to predict (via a
model) the BF correction for a red band. Alternatively, we could
certainly consider measuring flat-field statistics in $y$ band. 
We have not applied the same treatment to the $z$ band, where the
correction is much smaller than for $y$, and the need for reducing
the BF correction is considerably less obvious. We will however
apply the BF model for $z$ band in our final image processing. 

\section{Discussion}
\label{sec:discussion}

\subsection{Quality of the BF correction}
\begin{figure}[h]
  \begin{center}
        
    \includegraphics[width=\linewidth]{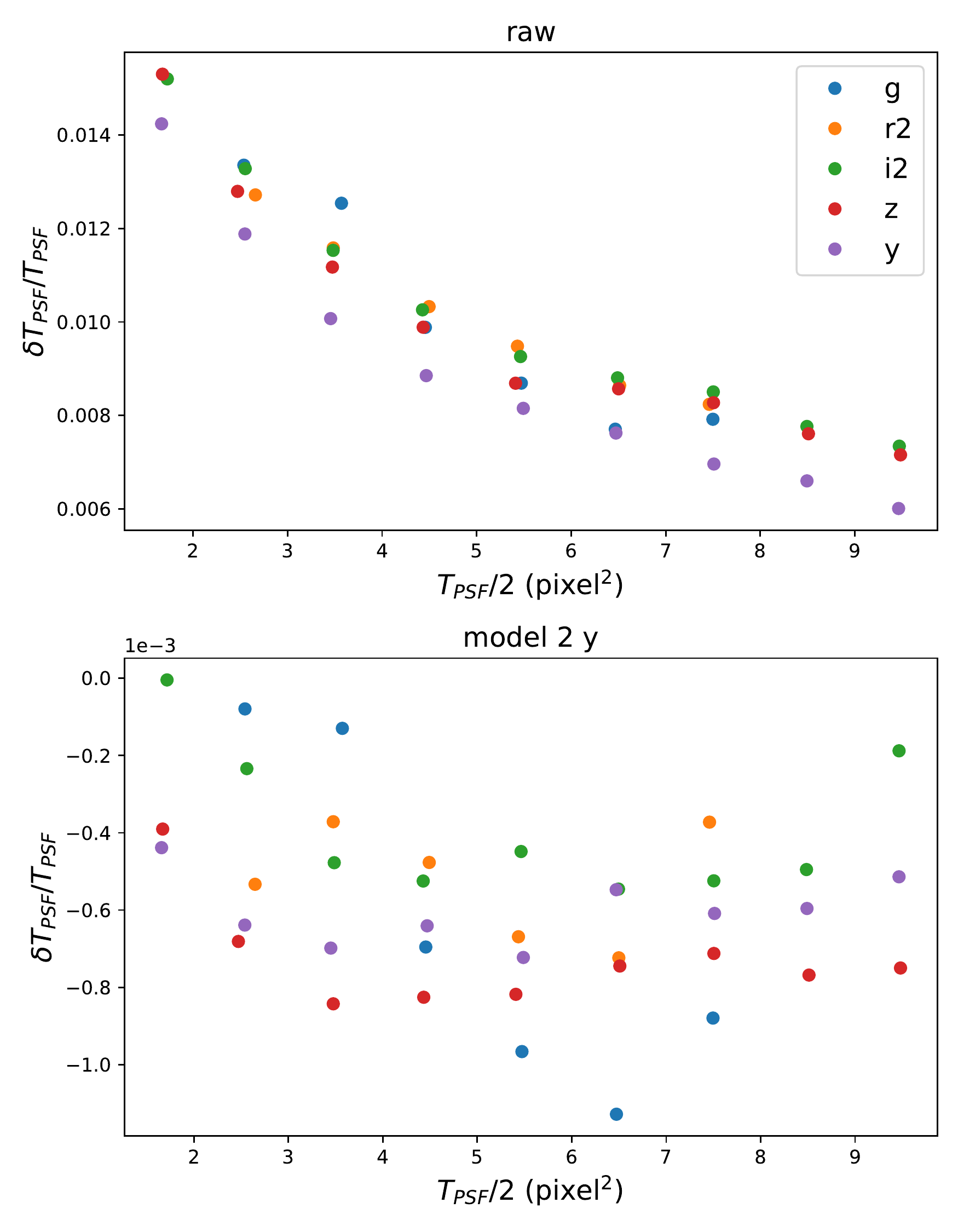}
    \caption{Relative change of the PSF size between an average PSF
      star and a faint object, as a function of $T_{PSF}/2$. For the
      PSF size extimator, we use as well the trace of the second
      moment matrix of the PSF. The top plot refers to the raw data, the bottom one to the data corrected by model 2 with the shortened drift paths in $y$ band.
      We assume that an average PSF star peaks at 1/3 of the saturation.
  \label{fig:plot_dt_over_t}
}
\end{center}
\end{figure}

\begin{figure}[h]
  \begin{center}
        
    \includegraphics[width=\linewidth]{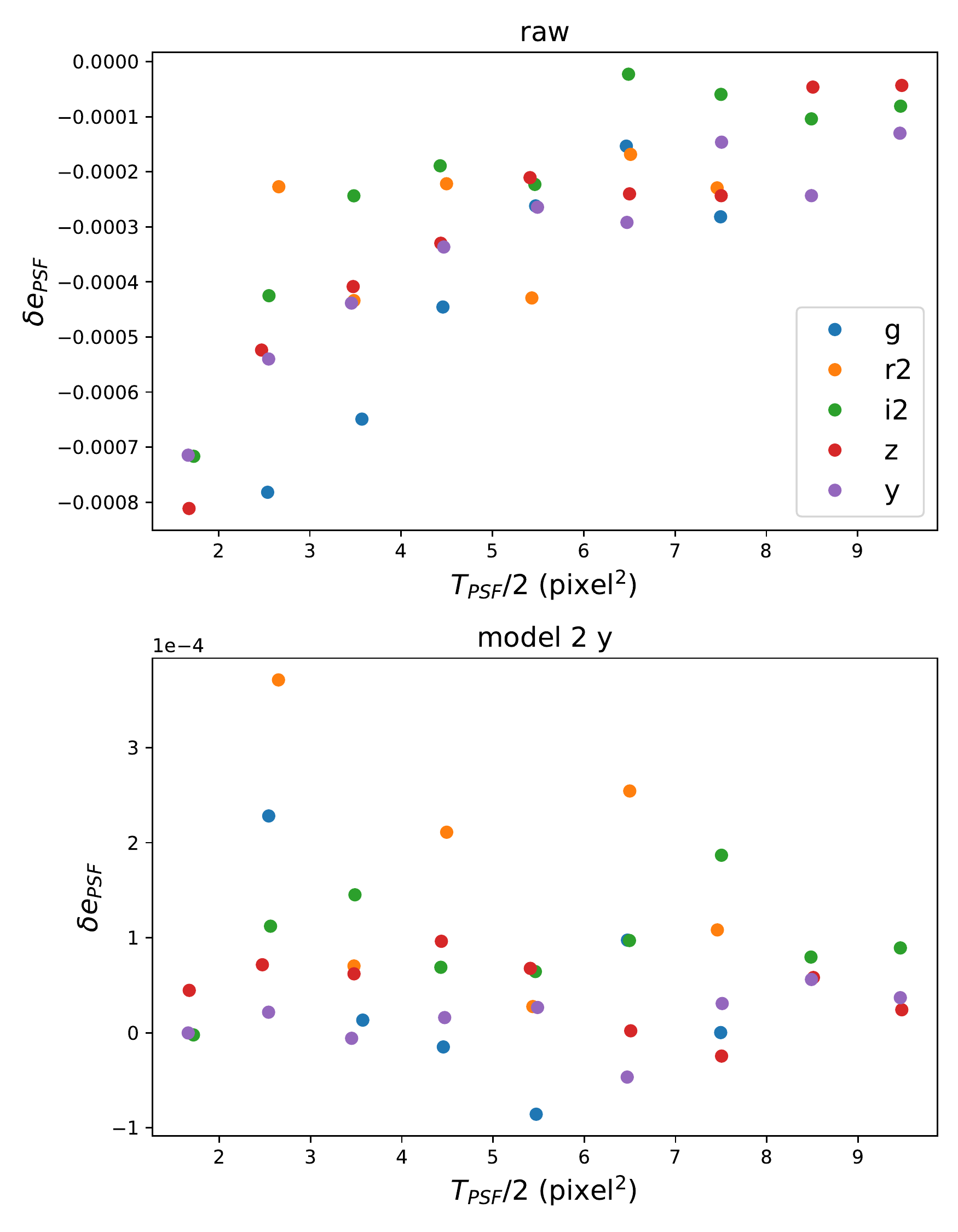}
    \caption{Change of the PSF ellipticity $e_{PSF} \equiv
      (M_{XX}-M_{YY})/T_{PSF}$ between an average PSF star and a faint
      object, as a function of $T_{PSF}/2$. The top plot refers to the
      raw data, the bottom one to the data corrected by model 2 with
      the shortened drift paths in $y$ band. We assume that an
      average PSF star peaks at 1/3 of the saturation.
  \label{fig:plot_dg1}
}
\end{center}
\end{figure}

\begin{figure}[h]
  \begin{center}
        
    \includegraphics[width=\linewidth]{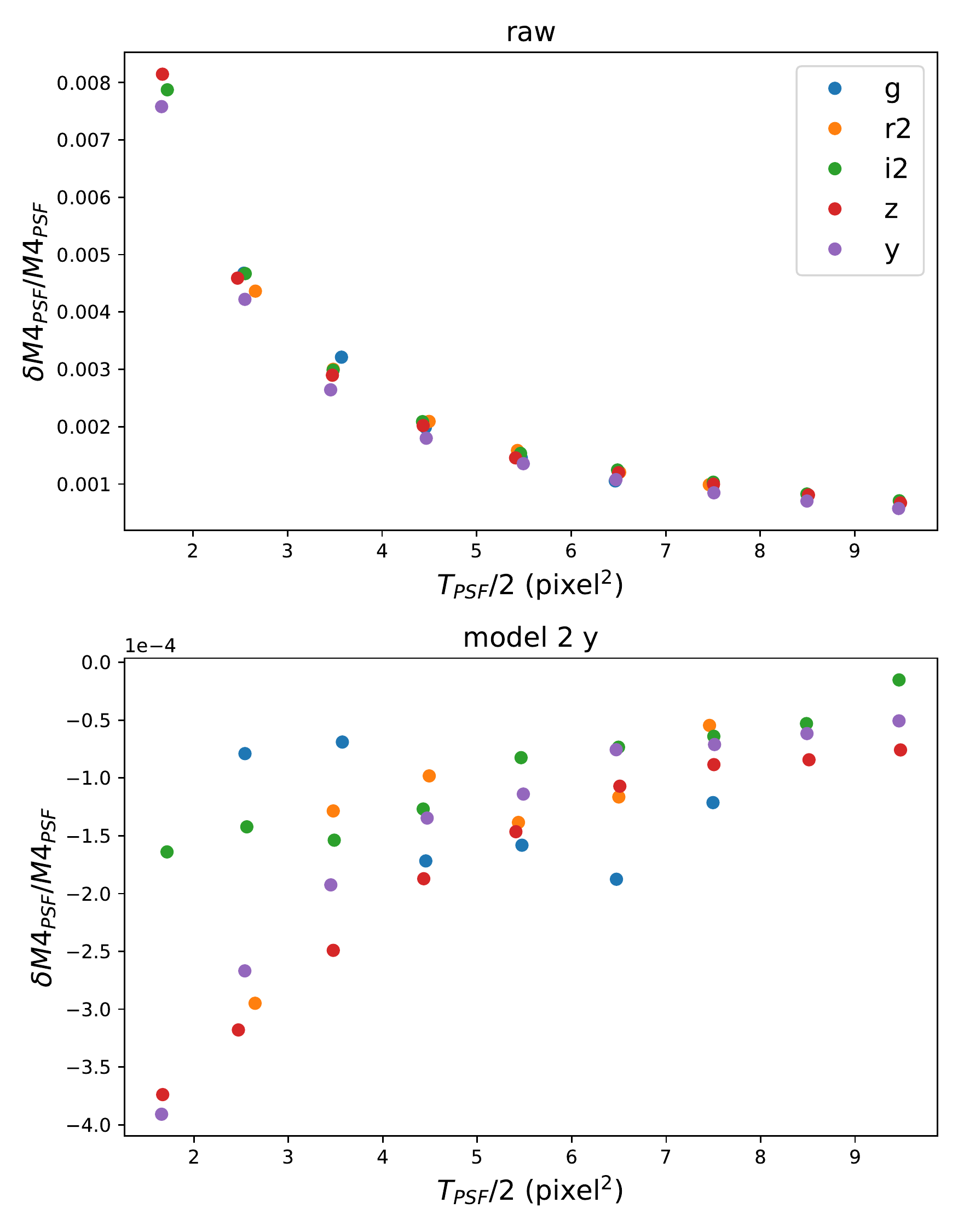}
    \caption{Relative change of the radial fourth moment as a
      function of image quality for all HSC bands, between a faint
      source and an average PSF star, before (top) and after (bottom)
      BF correction.  We assumed that an average PSF star peaks at 1/3
      of the saturation.
  \label{fig:plot_dmr4}
}
\end{center}
\end{figure}

The practical uses of PSF models crucially depend on the size of the
model capable of faithfully representing the actual PSF size for faint
objects. Two applications may come to mind: first, the measurement of
faint supernovae for cosmological applications where fluxes have to be
measured using PSF photometry and supernova fluxes are calibrated
against the ones of bright stars; second, the measurement of galaxy
shapes where one is interested in the intrinsic shape, namely, ``before''
it is smeared by the PSF. An inaccurate PSF size results in general in
a biased shape.  For both of these applications, the gauge is the
difference of the PSF model size to the real size of a faint object,
relative to the PSF size. \cite{DESC-sci-req} set the maximum
acceptable PSF size bias ($\delta T/T$) to $10^{-3}$ for the ten-year
Rubin/LSST survey for all sources of PSF size bias, where $T=M_{XX}+M_{YY}$
is the trace of the second moment matrix. So, in what follows, the PSF
``size'' is, in fact, the ``area''. Regarding PSF photometry, the relative
flux bias caused by PSF size bias reads $\delta f/f = 1/2 (\delta
T/T)$. Shear measurements also require an accurate estimation of the
PSF ellipticity. When surveys measure the same object with the same
orientation of the sensors, sensor-induced ellipticities are
transferred to shape measurements. For the {\it Euclid} mission
the requirement on ellipticities transferred to objects is
$5\ 10^{-5}$ of the rms (Table 1 of \citealt{Cropper13}). In the context of
the BF effect, we concentrate on the X/Y ellipticity, namely, $(M_{XX}-M_{YY})/T$. \cite{Cropper13} also provide a requirement for
the PSF size for {\it Euclid}, which reads $5\ 10^{-4}$ rms,
comparable to the Rubin $10^{-3}$ bound.

In Fig. \ref{fig:plot_dt_over_t}, we see that on corrected images,
the relative PSF size difference is below $10^{-3}$ in all bands at
all levels of image quality. In Fig. \ref{fig:plot_dg1}, we see that in
the corrected images, the residual PSF ellipticity
$\delta(M_{XX}-M_{YY})/T_{PSF}$ is mostly below $10^{-4}$ across the range of
image quality and meets the $5\ 10^{-5}$ r.m.s bound.
We thus note that the need to omit three nearby serial pixels when fitting
model 2 did not significantly degrade the quality of the BF correction
in one direction.
In Fig. \ref{fig:plot_dmr4}, we
display the slope of the radial fourth moment inaccuracies of the PSF,
which also induces adverse effects on shear measurements, with a
factor on the order of 1, as for the PSF size (see
\citealt{Zhang21}). The residuals are even smaller than for the
PSF size. We note that our residual trends after correction all
indicate an overcorrection of the BF effect and all this could thus be improved
by adjusting the overall scale of the correction. We also note that
these results were obtained in a blind way: we did not change
the procedure after having seen them.

\cite{Gruen-PACCD-15} evaluated their image correction for DECam in
ways similar to ours (their Fig. 12): their method overcorrects the
size of objects and leaves a negative BF slope, about -1/3 of the 
uncorrected one. This is certainly too large for a large-scale cosmic
shear survey. The corrected ellipticities seem significantly better
(and probably small enough for a DES-like survey) but they are weakly
affected by the BF effect in the test sample (as compared with the top
plot of our Fig. \ref{fig:plot_dg1}). \cite{Mandelbaum-HSC} assessed the
correction derived in C18 for HSC images: the (linear)
size of corrected stars varies with magnitude by about 0.2 \% over three
magnitudes (an eyeball estimation from their fig 6). Then, $T_{PSF}$
would evolve twice as much and this is small enough for the analysis of
the first year of the HSC survey.  Regarding the ellipticities, trends with regard to the flux
are not provided. For both DECam and HSC, the methods overcorrect
the BF trend and both assessments ignore chromatic contributions to
star size variations. If the color-flux correlation is similar to ours,
correcting sizes for color decreases the apparent BF slope, hence degrading the performance of the BF correction in both
instances.

We have insisted on the importance of higher order terms in
Eq. \ref{eq:C_ij} when analyzing the covariance curves. Those terms
result from the integration over time of Eq.
\ref{eq:delta_a}. Regarding the correction of the images, we did not carry out
integration over time, and we just considered that half of the
end-of-exposure image is a good approximation of the source of
electrostatic distortions, hence, neglecting second order effects which
are important for covariance curves. We anticipate that neglected second
order effects in image correction should manifest themselves as
structured residuals to the moments versus $f_{max}$ linear fits. We analyzed these
residuals in band and image quality bins, summing different bands at
the same image quality, and we could not find any hint of departure from
linearity. This is fortunate since accounting for next to leading order
effects in image correction is more difficult than for the shape of
covariance curves.

\subsection{Possible further developments}
While the obtained performance of the BF correction seems sufficient,
we may consider potential avenues for improvements. First, in our analysis, the overall normalization of the BF
correction model primarily depends on $a_{01}$, which is the best
measured area coefficient actually used in the electrostatic fit (the
model 2 fit). This coefficient is slightly biased by an erroneous value for $a_{00}$
through the $\mu^3$ terms in the covariance fit to a level compatible
with the small over-correction we are facing. It may then seem
legitimate to actually tune the overall normalization of
the correction in order to bring the average BF slope to 0,
at least for the best observing conditions.

Second, the BF correction we implemented is the outcome of an analysis
where three area coefficients of the BF effect (the three first serial
measurements, including the largest coefficient) had to be ignored
because they could not be accommodated by an electrostatic model. Those
coefficients had to be derived from the measurements of other
area coefficients through an electrostatic modeling.
We might then anticipate that for a camera that does not
suffer from this noise bias (which is first manifested by a
violation of the sum rule), a better correction model could be
constructed. In particular, an analysis of test data of the Vera Rubin
Observatory camera indicates that the sum rule is satisfied on the
integrated instrument.

One appealing -- and very different -- approach could be to determine the
electrostatic model from the science data by fitting the pixel
distortion pattern that makes the PSF homothetic. The electrostatic
modeling could probably be computed sufficiently rapidly to be
inserted into a PSF modeling fitting loop. The concerns about
the precision measurement of the two-point function of flat
fields would become pointless.

\subsection{Computer codes}
Our code is split in three different parts: code to measure
covariances and fit the covariance curves, code to perform the
electrostatic fit, and code to process the images and in particular
corrects those for the BF effect. Our public
repository\footnote{\href{https://gitlab.in2p3.fr/astier/bfptc}{https://gitlab.in2p3.fr/astier/bfptc}}
contains the python code for the two first steps. The code that
measures covariances did not evolve significantly since it was
developed for A19. We publish here the electrostatic modeling code
for the first time. Our image correction code is fairly straightforward
and would probably run much faster with 
the convolution in Eq. \ref{eq:boundary_shifts} computed in Fourier space. The parts of the analysis that are not in the repository can be made
available upon request.

\begin{acknowledgements}
  We are indebt to the Subaru Telescope technical staff that very
  efficiently operates the observatory and its instruments. HSC images
  are made available on the SMOKA
  server\footnote{\href{https://smoka.nao.ac.jp/index.jsp}{https://smoka.nao.ac.jp/index.jsp}}, that we have
  used extensively. We perform all our reductions and store our
  results at the Centre de Calcul de
  l'IN2P3\footnote{\href{https://cc.in2p3.fr}{https://cc.in2p3.fr}}, a computing facility of CNRS.
  This work benefited from useful discussions with N. Suzuky (LBL) and
  N. Yasuda (IPMU), and our colleagues from the Paris team. The manuscript
  eventually benefited from excellent suggestions from our referee. 
\end{acknowledgements}

\def\aap{A\&A}
\def\nat{Nature}
\def\mnras{MNRAS}
\def\memras{MmRAS}
\def\aapr{The Astron. and Astrop. Rev.}
\def\prd{Phys. Rev. D}
\def\sovast{Soviet Astronomy}
\def\jcap{J. Cosm. Astropart. P.}
\def\aj {Astron. Journ.}

\bibliography{biblio}

\end{document}